\documentclass{article}
\usepackage{algorithm,algorithmic}
\usepackage{amsmath,amssymb,amsthm,hyperref}
\usepackage[titletoc]{appendix}
\usepackage{relsize}
\usepackage{authblk}
\usepackage{subfigure}
\usepackage{verbatim}
\usepackage{setspace}
\usepackage[margin=1.0in]{geometry}
\usepackage{tikz}
\usepackage{multicol}
\usepackage{adjustbox}
\usepackage{thmtools}
\usepackage{thm-restate}
\usetikzlibrary{shapes,arrows}

\newif\ifnotes\notestrue

\ifnotes

\newcommand{\anote}[1]{\textcolor{red}{{\bf (Alex:} {#1}{\bf ) }} \marginpar{\tiny\bf
             \begin{minipage}[t]{0.5in}
               \raggedright ??
            \end{minipage}}}
\newcommand{\nnote}[1]{\textcolor{red}{{\bf (Naman:} {#1}{\bf ) }} \marginpar{\tiny\bf
             \begin{minipage}[t]{0.5in}
               \raggedright ??
            \end{minipage}}}            
\newcommand{\vnote}[1]{\textcolor{blue}{{\bf (Vivek:} {#1}{\bf ) }} \marginpar{\tiny\bf
             \begin{minipage}[t]{0.5in}
               \raggedright ??
                \end{minipage}}}  
\newcommand{\knote}[1]{\textcolor{red}{{\bf (Karthik:} {#1}{\bf ) }} \marginpar{\tiny\bf
             \begin{minipage}[t]{0.5in}
               \raggedright ??
            \end{minipage}}}
\else
\newcommand{\anote}[1]{}
\newcommand{\nnote}[1]{}
\newcommand{\vnote}[1]{}
\newcommand{\knote}[1]{}
\fi

\newtheorem{fact}{Fact}
\newtheorem{theorem}{Theorem}
\newtheorem{lemma}[theorem]{Lemma}
\newtheorem{claim}[theorem]{Claim}
\newtheorem{definition}[theorem]{Definition}
\newtheorem{corollary}[theorem]{Corollary}
\newcommand{\LN}{\lambda_{new}}
\newcommand{\N}{\mathbb{N}}
\newcommand{\C}{\mathbb{C}}
\newcommand{\R}{\mathbb{R}}
\newcommand{\Z}{\mathbb{Z}}
\newcommand{\Oh}{\mathcal{O}}

\newcommand{\V}{\mathcal{V}}

\begin{document}
\title{On the Expansion of Group--based Lifts}
\author{Naman Agarwal\footnote{namana@cs.princeton.edu, Princeton University} , 
Karthekeyan Chandrasekaran\footnote{karthe@illinois.edu, University of Illinois Urbana-Champaign} , 
Alexandra Kolla\footnote{akolla@illinois.edu, University of Illinois Urbana-Champaign} , 
Vivek Madan\footnote{vmadan2@illinois.edu, University of Illinois Urbana-Champaign}}
\date{}
\maketitle


\thispagestyle{empty}
\begin{singlespace}
\section*{\centering{Abstract}}
A $k$-lift of an $n$-vertex base graph $G$ is a graph $H$ on $n\times k$ vertices, where each vertex $v$ of $G$ is replaced by $k$ vertices $v_1,\cdots{},v_k$ and each edge $(u,v)$ in $G$ is replaced by a matching representing a bijection $\pi_{uv}$ so that the edges of $H$ are of the form $(u_i,v_{\pi_{uv}(i)})$. Lifts have been studied as a means to efficiently construct expanders. In this work, we study lifts obtained from \emph{groups and group actions}. We derive the spectrum of such lifts via the representation theory principles of the underlying group. Our main results are: 
\begin{enumerate}
\item 
There is a constant $c_1$ such that for every $k\geq 2^{c_1nd}$, there \emph{does not} exist an abelian $k$-lift $H$ of any $n$-vertex $d$-regular base graph with $H$ being \emph{almost Ramanujan} (nontrivial eigenvalues of the adjacency matrix at most $\Oh(\sqrt{d})$ in magnitude). This can be viewed as an analogue of the well-known no-expansion result for abelian Cayley graphs. 
\item 
A uniform random lift in a cyclic group of order $k$ of any $n$-vertex $d$-regular base graph $G$, with the nontrivial eigenvalues of the adjacency matrix of $G$ 
bounded by $\lambda$ in magnitude, has the new nontrivial eigenvalues also bounded by $\lambda+\Oh(\sqrt{d})$ in magnitude with probability $1-ke^{-\Omega(n/d^2)}$. In particular, there is a constant $c_2$ such that for every $k\leq 2^{c_2n/d^2}$, there exists a lift $H$ of every Ramanujan graph in a cyclic group of order $k$ with $H$ being almost Ramanujan. We use this fact to design a quasi-polynomial time algorithm to construct almost Ramanujan expanders deterministically.
\end{enumerate}
The existence of expanding lifts in cyclic groups of order $k=2^{\Oh(n/d^2)}$ can be viewed as a lower bound on the order $k_0$ of the largest abelian group that produces expanding lifts. Our two results show that the lower bound closely matches the upper bound for $k_0$ (upto a factor of $d^3$ in the exponent), thus suggesting a threshold phenomenon.

\clearpage
\setcounter{page}{1}
\section{Introduction}

Expander graphs have spawned research in pure and applied mathematics during the last several years, with several applications to multiple fields including complexity theory, the design of robust computer networks, the design of error-correcting codes, de-randomization of randomized algorithms, compressed sensing and the study of metric embeddings. For a comprehensive survey of expander graphs see \cite{Sar04, HLW06}.

Informally, an expander is a graph where every small subset of the vertices has a relatively
large edge boundary. Most applications are concerned with sparse $d$-regular graphs $G$, where the largest eigenvalue of the adjacency matrix $A_G$ is $d$. In case of a bipartite graph, the largest and smallest eigenvalues of $A_G$ are $d$ and $-d$, which are referred to as trivial eigenvalues. The expansion of the graph is related to the difference between $d$ and $\lambda$, the first largest (in absolute value) non-trivial eigenvalue of $A_G$. Roughly, the smaller $\lambda$ is, the better the graph expansion. The Alon-Boppana bound (\cite{Nil91}) states that $\lambda \geq 2\sqrt{d-1}-o(1)$ for non-bipartite graphs, thus graphs with $\lambda \leq 2\sqrt{d-1}$ are optimal expanders and are called Ramanujan.

 A simple probabilistic argument can show the existence of infinite families of expander graphs \cite{Pin73}. However, constructing such infinite families explicitly has proven to be a challenging and important task. It is easy to construct Ramanujan graphs with a small number of vertices: $d$-regular complete graphs and complete bipartite graphs are Ramanujan. The challenge is to construct an infinite family of $d$-regular graphs that are all Ramanujan, which was first achieved by  Lubotzky, Phillips and Sarnak \cite{LPS88} and Margulis \cite{Mar88}. They built Ramanujan graphs from Cayley graphs. All of their graphs are regular, have degrees $p+1$ where $p$ is a prime, and their proofs rely on deep number theoretic facts. In two recent breakthrough papers, Marcus, Spielman, and Srivastava showed the existence of bipartite Ramanujan graphs of all degrees \cite{MSS13, MSS15}. However their results do not provide an efficient algorithm to construct those graphs. A striking result of Friedman \cite{Fri08} and a slightly weaker but more general result of Puder \cite{Pud13}, shows that almost every $d$-regular graph on n vertices is very close to being Ramanujan i.e. for every $\epsilon>0$, asymptotically almost surely,
$\lambda <  2	\sqrt{d-1} +\epsilon$. It is still unknown whether the event that a random $d$-regular graph is exactly Ramanujan happens with constant probability. Despite the large body of work on the topic, all attempts to efficiently construct large Ramanujan expanders of any given degree have failed, and exhibiting such constructions remains an intriguing open problem.

A combinatorial approach to constructing expanders, initiated by Friedman \cite{Fri03}, is
to prove that one may obtain new (larger) Ramanujan graphs from smaller ones. In
this approach, one starts with a base graph $G$ which one ``lifts'' to obtain a
larger graph $H$. 
More concretely, a $k$-lift of an $n$-vertex base-graph $G$ is a graph $H$ on $k\times n$ vertices , where each vertex $u$ of $G$ is replaced by $k$ vertices $u_1,\cdots{},u_k$ and each edge $(u,v)$ in $G$ is replaced by a matching between $u_1,\cdots{},u_k$ and $v_1,\cdots{},v_k$. In other words, for each edge $(u,v)$ of $G$ there is a permutation $\pi_{uv}$ of $k$ elements so that the corresponding $k$ edges of $H$ are of the form $(u_i,v_{\pi_{uv}(i)})$. The graph $H$ is a (uniformly) \textit{random} lift of $G$ if for every edge $(u,v)$ the bijection $\pi_{uv}$ is chosen uniformly and independently at random from the set of permutations of $k$ elements, $S_k$.

Since we are focusing on Ramanujan graphs, we will restrict our attention to lifts of $d$-regular graphs. It is easy to see that any lift $H$ of a $d$-regular base-graph $G$ is itself $d$-regular and inherits all the eigenvalues of $G$ (which, hereafter we refer to as ``old'' eigenvalues, whereas the rest of the eigenvalues are referred to as ``new'' eigenvalues). In order to use lifts for building expanders, it is necessary that the lift would also inherit the expansion properties of its base graph. One hopes that a random lift of a Ramanujan graph will also be (almost) Ramanujan with high probability. 

Friedman \cite{Fri03} first studied the eigenvalues of random $k$-lifts of regular graphs and proved that every new eigenvalue of $H$ is $\Oh(d^{3/4})$ with high probability. He conjectured a bound of $2\sqrt{d-1}+ o(1)$, which would be tight (see, e.g. \cite{LG}). Linial and Puder \cite{LP10} improved Friedman's bound to $\Oh(d^{2/3})$. Lubetzky, Sudakov and Vu \cite{LSV11} showed that the absolute value of every nontrivial eigenvalue of the lift is $ \Oh(\lambda  \log d )$, where $\lambda$ is the second largest (in absolute value) eigenvalue of the base graph, improving on the previous results when $G$ is significantly expanding. Adarrio-Berry and Griffiths \cite{ABG10} further improved the bounds above by showing that every new eigenvalue of $H$ is $\Oh(\sqrt{d})$, and Puder \cite{Pud13} proved the nearly-optimal bound of $2\sqrt{d-1}+1$. All those results hold with probability tending to $1$ as $k\rightarrow \infty$, thus the order $k$ of the lift in question needs to be large. Nearly no results were known in the regime where $k$ is bounded with respect to the number of nodes $n$ of the graph. A ``relativized'' version of the Alon-Boppana Conjecture regarding lower-bounding the new eigenvalues of lifts was also recently shown in\cite{FK14} and \cite{Bon15}.

 Bilu and Linial \cite{BL06} were the first to study $k$-lifts of graphs with bounded $k$, and suggested constructing Ramanujan graphs through a sequence of 2-lifts of a base graph: start with a small $d$-regular Ramanujan graph on some finite number of nodes (e.g. $K_{d+1}$). Every time the 2-lift operation is performed, the size of the graph doubles. If there is a way to preserve expansion after lifting, then repeating this operation will give large good expanders of the same bounded degree $d$. The authors in \cite{BL06} showed that if the starting graph $G$ is significantly expanding so that $\lambda(G)=\Oh (\sqrt{d\log d})$, then there exists a random 2-lift of $G$ that has all its new eigenvalues upper-bounded in absolute value by $\Oh(\sqrt{d\log^3 d})$. In the recent breakthrough work of Marcus, Spielman and Srivastava \cite{MSS13}, the authors showed that for every bipartite graph $G$, there exists a 2-lift of $G$, such that the new eigenvalues achieve the Ramanujan bound of $2\sqrt{d-1}$, but their result still does not provide any efficient algorithm to find such lifts. 

\subsection{Our Results}
In this work, we study lifts as a means to efficiently construct almost Ramanujan expanders of all degrees. We derive these lifts from groups. This is a natural generalization of Cayley graphs.
\begin{definition}[$\Gamma$-lift]
Let $\Gamma$ be a group of order $k$ with $\cdot$ denoting the group operation. A $\Gamma$-lift of an $n$-vertex base graph $G=(V,E)$ is a graph $H=(V \times \Gamma, E')$ obtained as follows: it has $k\times n$ vertices, where each vertex $u$ of $G$ is replaced by $k$ vertices $\{u\}\times \Gamma$. For each edge $(u,v)$ of $G$, we choose an element $g_{u,v}\in \Gamma$ and replace that edge by a perfect matching between $\{u\}\times \Gamma$ and $\{v\}\times \Gamma$ that is given by the edges $(u_i,v_j)$ for which $g_{u,v}\cdot i = j$. 

We denote $|\Gamma|=k$ to be the order of the lift. We refer to $\Gamma$-lifts obtained using $\Gamma=\Z/k\Z$, the additive group of integers modulo $k$, as shift $k$-lifts. Since every cyclic group of order $k$ is isomorphic to $\Z/k\Z$, we have that $\Gamma$-lifts are shift $k$-lifts whenever $\Gamma$ is a cyclic group. 

\end{definition}

A tight connection between the spectrum of $\Gamma$-lifts and the representation theory of the underlying group $\Gamma$ is known \cite{MS95,Oldrepresentation}. This connection tells us that the lift graph incurs the eigenvalues of the base graph, while its new eigenvalues are the union of eigenvalues of a collection of matrices arising from the irreducible representations of the group and the group elements assigned to the edges. This connection has been recently used in \cite{HPS15} in the context of expansion of lifts, aiming to generalize the results in \cite{MSS15}. In order to understand the expansion properties of the lifts, we focus on the new eigenvalues of the lifted graph. We address the expansion of $\Gamma$-lifts obtained from cyclic groups and abelian groups. 

We present a high probability bound on the expansion of random shift $k$-lifts for bounded $k$. 


\begin{restatable}{theorem}{thmShiftk}\label{thm:main2}
Let $G$ be a $d$-regular $n$-vertex graph with non-trivial eigenvalues at most $\lambda$ in absolute value where $\lambda\ge \sqrt{d}$, $2 \le d \leq \sqrt{n/(3\ln n)}$, and $H$ be a random shift $k$-lift of $G$. Let $\lambda_{new}$ be the largest new eigenvalue of $H$ in magnitude. Then $$\lambda_{new} = \Oh(\lambda)$$
with probability $1-k\cdot e^{-\Omega(n/d^2)}$. Moreover, if $G$ is
moderately expanding such that $\lambda \leq d/\log{d}$, then
$$\lambda_{new}-\lambda=\Oh(\sqrt{d})$$ with probability $1-k\cdot e^{-\Omega(n/d^2)}$.
\end{restatable}

In particular, if we start with $G$ being a Ramanujan expander, then w.h.p. a random shift $k$-lift will be almost Ramanujan, having all its new eigenvalues bounded by $\Oh(\sqrt{d})$.

\paragraph{Remark 1.}
In contrast to the case of lifts of order $k\rightarrow \infty$, the dependency on $\lambda$ is necessary for bounded $k$. This has previously been observed by the authors in \cite{BL06} who gave the following example: Let $G$ be a disconnected graph on $n$ vertices that consists of $n/(d+1)$ copies of $K_{d+1}$, and let $H$ be a random 2-lift of $G$. Then the largest non-trivial eigenvalue of $G$ is $\lambda=d$ and it can be shown that with high probability, $\lambda_{new}=\lambda=d$. Therefore, our eigenvalue bounds are nearly tight.\\

We state Theorem \ref{thm:main2} specialized for the case of $2$-lifts next. 

\begin{restatable}{corollary}{thm2Lifts}\label{thm:main1}
Let $G$ be a $d$-regular $n$-vertex graph with non-trivial eigenvalues at most $\lambda$ in absolute value where $\lambda\ge \sqrt{d}$, $2 \le d \leq \sqrt{n/(3\ln n)}$, and $H$ be a uniformly random $2$-lift of $G$. Let $\lambda_{new}$ be the largest new eigenvalue of $H$ in magnitude. Then $$\lambda_{new} = \Oh(\lambda)$$
with probability $1-e^{-\Omega(n/d^2)}$. Moreover, if $G$ is moderately
expanding such that $\lambda \leq d/\log d$, then $$\lambda_{new}-\lambda =\Oh(\sqrt{d})$$ with probability $1-e^{-\Omega(n/d^2)}$.
\end{restatable}

\paragraph{Remark 2.}
Our result for $2$-lifts improves upon the $\log d$ factor present in the result of Bilu-Linial \cite{BL06}. This factor arises in their analysis due to the use of the converse of the Expander Mixing Lemma (EML) along with an $\epsilon$-net style argument. The converse EML is provably tight, so straightforward use of the converse EML will indeed incur the $\log d$ factor. We are able to improve the eigenvalue bound by performing a deeper analysis of the $\epsilon$-net argument, avoiding direct use of the converse EML.\\



Lifts based on groups immediately suggest an algorithm towards building $d$-regular $n$-vertex Ramanujan expanders. In order to describe this algorithm, we first describe the brute-force algorithm that follows from the existential result of \cite{MSS13}. The idea is to start with the complete bipartite graph $K_{d,d}$ and lift the graph $\log_2(n/2d)$ times. At each stage, we brute force search over the space of all possible 2-lifts and pick the best (most expanding) one. However, since a graph $(V, E)$ has $2^{|E|}$ possible 2-lifts, it follows that the final lift will be chosen from among $2^{nd/4}$ possible 2-lifts, which means that the brute force algorithm will run in time exponential in $nd$. 

Next, suppose that for every $k\ge 2$, we are guaranteed the existence of a group $\Gamma$ of order $k$ such that for every base graph there exists a $\Gamma$-lift that has all its new eigenvalues at most $2\sqrt{d-1}$ in absolute value. For example, Chandrasekaran-Velingker \cite{CV15} suggest the possibility that for every $k$ and for every base graph, there exists a shift $k$-lift that has all new eigenvalues with absolute value at most $2\sqrt{d-1}$. Then a brute force algorithm similar to the one above, would perform only one lift operation of the base graph $K_{d,d}$ to create a $\Gamma$-lift with $n=2dk$ vertices. This algorithm would only have to choose the best among $k^{d^2}$ possibilities ($k$ different choices of group element per edge of the base graph), which is polynomial in $n$, the size of the constructed graph. Here we have assumed that $d$ is a constant. This motivates the following question: what is the largest possible group $\Gamma$ that might produce expanding $\Gamma$-lifts? 
Our next result rules out the existence of large abelian groups that might lead to (even slightly) expanding lifts.

\begin{restatable}{theorem}{thmAbelian}
\label{thm:abelian_lift_not_exists}
For every $n$-vertex $d$-regular graph $G$, $\epsilon\in (0,1)$, and abelian group $\Gamma$ of size at least
\[
k=exp\left(\frac{nd\log\frac{1}{\epsilon}+\log{n}}{\log{\frac{1}{e\epsilon}}}\right),
\]
all $\Gamma$-lifts $H$ of $G$ have $\lambda(H)$ at least $\epsilon d$. In particular, when $k=2^{\Omega(nd)}$, there is no $\Gamma$-lift $H$ of any $n$-vertex $d$-regular graph $G$ with $\lambda(H)=\Oh(\sqrt{d})$ whenever $\Gamma$ is an abelian group of order $k$.
\end{restatable}

We note that Theorem \ref{thm:abelian_lift_not_exists} shows that one cannot expect to have arbitrarily large abelian groups with expanding lifts as suggested by \cite{CV15}. 

\paragraph{Remark 3.}
The first and only known efficient constructions of Ramanujan expanders are Cayley graphs of certain groups \cite{LPS88}. 
We observe that a Cayley graph for a group $\Gamma$ with generator set $S$ can be obtained as a $\Gamma$-lift of the bouquet graph (a graph that consists of one vertex with multiple self loops) \cite{Makelov}. Our no-expansion result for abelian groups complements the known result on no-expansion of abelian Cayley graphs \cite{Abelian}.



\paragraph{Remark 4.}
Our Theorems \ref{thm:abelian_lift_not_exists} and \ref{thm:main2} can be viewed as lower and upper bounds on the largest order $k_0$ of an abelian group $\Gamma$ such that for every $n$-vertex graph, there exists a $\Gamma$-lift for which all the new eigenvalues are small. On the one hand, Theorem \ref{thm:main2} shows that, for $k= 2^{\Oh(n/d^2)}$, most of the shift $k$-lifts of a Ramanujan graph have their new eigenvalues upper-bounded by $\Oh(\sqrt{d})$. On the other hand, Theorem \ref{thm:abelian_lift_not_exists} shows that for $k=2^{\Omega(nd)}$, there is no shift $k$-lift that achieves such expansion guarantees. This suggests a threshold behavior for $k_0$. 
\\
%


Moreover, Theorem \ref{thm:main2} leads to a deterministic quasi-polynomial time algorithm for constructing almost Ramanujan (with $\lambda = \Oh(\sqrt{d})$) families of graphs. 
\begin{restatable}{theorem}{thmAlgo}\label{thm:algo}
There exists an algorithm to construct a $d$-regular $n$-vertex graph $G$ such that $\lambda(G)=\Oh(\sqrt{d})$ in $2^{\Oh(d^4\log^2 n)}$ time.
\end{restatable}

\begin{algorithm}[htb]\label{algo}
	\caption{Quasi-polynomial time algorithm to construct expanders of arbitrary size $n$}
	\label{alg:quassi-poly-expander}
	\begin{algorithmic}[1]
		\STATE Pick an $r$ such that $2^{ c r/d^2}\cdot r=n$, for a constant $c$ given by Theorem \ref{thm:main2}. Do an exhaustive search to find a $d$-regular graph $G'$ on $r$ vertices with $\lambda = \Oh(\sqrt{d})$. 
		\STATE For $k = 2^{c r/d^2}$, do an exhaustive search to find a shift $k$-lift $G$ of $G'$ with minimum $\lambda(G)$.
	\end{algorithmic}
\end{algorithm}

\begin{proof}
We use Algorithm 1. We note that the choice of $r$ in the first step ensures that $r=\Oh(d^2 \log n)$. 
By Theorem \ref{thm:main2}, there exists a lift $G$ of $G'$ such that $\lambda(G)=\Oh(\sqrt{d})$. Thus, the exhaustive search in the second step gives a graph $G$ with $\lambda(G)=\Oh(\sqrt{d})$.

For the running time, we note that the first step can be implemented to run in time $2^{\Oh(r^2)}=2^{\Oh(d^4\log^2{n})}$. To bound the running time of the second step, we observe that for each edge in $G'$, there are $k$ possible choices. Therefore the total search space is at most $k^{rd/2}=2^{cr^2/2d}=2^{\Oh(d^3 \log^2{n})}$ and for each $k$-lift, it takes $poly(n)$ time to compute $\lambda(G)$. Thus, the overall running time of the algorithm is $2^{\Oh(d^4 \log^2{n})}$.

\end{proof}



\noindent {\bf Organization.}
We give some preliminary definitions, notations, facts and lemmas in Section \ref{sec:prelims}. We recall the tight connection between the spectrum of $\Gamma$-lifts and the representation of the group $\Gamma$ in Section \ref{sec:spectrum-of-lifts}. We prove Theorem \ref{thm:abelian_lift_not_exists} in Section \ref{sec:abelian-lifts}. 
We illustrate the techniques behind proving Theorem \ref{thm:main2} by presenting and proving 
a slightly weaker version of Theorem \ref{thm:main2} (see Theorem \ref{main_theorem_easy}) in Section \ref{sec:proofs_short}. 
We prove the concentration inequality (Lemma \ref{main_lemma}) needed for the weaker version in Section \ref{sec:proof_lemma}. 
We use a stronger version of the concentration inequality and prove Theorem \ref{thm:main2} in Section \ref{proof-of-random-shift-lifts} of the appendix.


\section{Preliminaries}\label{sec:prelims}
In this section, we set the notation and present the needed combinatorial inequalities and facts. 
\paragraph{Notations.}
Let $G=(V,E)$ be a $d$-regular graph with $n$ vertices. 
If $G$ is $d$-regular bipartite, we will assume that the bipartition of the vertex set is given by $(\{1,\ldots,n/2\},\{n/2+1,\ldots, n\})$.
Let $A$ be the adjacency matrix of $G$, and $\lambda_1 \geq \lambda_2 \geq \ldots
\lambda_n$ be its $n$ eigenvalues. Since $A$ is a real symmetric matrix, its eigenvalues are
also real. For a $d$-regular graph $G$, it is well-known that
$\lambda_1 = d$. 
If $G$ is bipartite, then $\lambda_n=-d$ and we define $\lambda_G :=\max \limits_{i:[2,n-1]}|\lambda_i|$. 
If $G$ is non-bipartite, we define $\lambda_G := \max \limits_{i:[2,n]}|\lambda_i|$. Thus, $\lambda_G$ denotes the maximum non-trivial eigenvalue of $G$. When $G$ is clear from the context, we will drop the subscript and simply write $\lambda$. 
For subsets $S,T \subseteq V$, let $E(S,T)$ be the number of edges $uv\in E$ with $u\in S$ and $v\in T$. 
We denote the largest eigenvalue of a matrix $M$ by $\|M\|$ and the support of a vector $x$ by $S(x)$. 
We define $\log()$ to be the $\log$ function with base 2. We represent $e^x$ by $exp(x)$.
Given a vector $x \in \{0,\pm 1/2,\pm 1/4 \ldots \}$ we define the \emph{diadic
decomposition} of $x$ as the set $\{2^{-i}u_i\}$ where each $u_i$ is defined as
\[
[u_i]_j = \begin{cases}
	1 & \text{if  } x_j = 2^{-i},\\
	-1 & \text{if  } x_j = -2^{-i},\\
    0 & \text{otherwise}.
\end{cases}
\]

We will use the Hoeffding inequality for concentration bounds. 
\begin{theorem}[Hoeffding inequality]\label{thm:hoeffding_inequality}
Let $X_1,\dots, X_n$ be independent random variables such that $X_i$ is strictly bounded within the interval $[a_i,b_i]$, then 
\begin{equation*}
\mathbb{P}\left(\left|\sum_{i=1}^n X_i - \sum_{i=1}^n E[X_i]\right| \geq t\right) \leq 2e^{-\frac{2t^2}{\sum_{i=1}^n (b_i-a_i)^2}}.
\end{equation*}
\end{theorem}

We use the following combinatorial identity. 

\begin{lemma}[Discretization Lemma]
\label{DiscretizationLemma}
Let $M\in \R^{n\times n}$ be a matrix with diagonal entries being $0$. 
\begin{enumerate}
\item For every $x\in \R ^n$ with $||x||_\infty \leq 1/2$ there exists $y \in \{\pm 1/2,\pm 1/4, \ldots \}^n$ such
that $|x^TMx|\leq |y^TMy|$ and $\|y\|^2 \leq 4\|x\|^2$. Moreover, each entry of
$x$ between $2^{-i}$ and $2^{-i-1}$ is rounded to either $2^{-i}$ or
$2^{-i-1}$ and between $-2^{-i}$ and $-2^{-i-1}$ is rounded to either $-2^{-i}$ or $-2^{-i-1}$ in $y$.

\item 
Similarly, for every $x_1,x_2\in \R^n$ with $||x_1||_\infty, ||x_2||_\infty \leq 1/2$,
there exist $y_1, y_2 \in \{\pm 1/2, \pm 1/4, \ldots \}^n$ such that $|x_1^T M
x_2| \leq |y_1^T M y_2|$, $\|y_1\|^2 \leq 4\|x_1\|^2$, $\|y_2\|^2 \leq
4\|x_2\|^2$ and each entry of $x_1,x_2$ between $2^{-i}$ and $2^{-i-1}$ is
rounded to either $2^{-i}$ or $2^{-i-1}$ and between $-2^{-i}$ and $-2^{-i-1}$ is rounded to either $-2^{-i}$ or $-2^{-i-1}$ in $y_1, y_2$.
\end{enumerate}
\end{lemma}
\begin{proof}
To obtain such a vector $y$ we take a vector $x$ and
round its coordinates independently with the following probabilistic rule. Let
$x_i = \pm(1+\delta_i)2^{-i}$ be the $i^{th}$ coordinate of $x$. We round $x_i$ to
$sign(x_i)\cdot 2^{-i+1}$ with probability $\delta_i$ and $sign(x_i)\cdot 2^{-i}$ with
probability $1 - \delta_i$. Let the rounded vector be $x'$. We note that $E[x_i']
= x_i$. Since each coordinate is rounded independently and the diagonal
entries of M are 0, we get that $E[x'^TMx'] = x^TMx$. This implies that there exists
a vector $y \in \{\pm 1/2,\pm 1/4, \ldots \}^n$  that can be generated by this rounding
such that $|x^TMx|\leq |y^TMy|$. Also it is easy to see that $\|y\|^2 \leq
4\|x\|^2$ and by definition every coordinate in $y$ with value between $\pm2^{-i}$ and
$\pm2^{-i-1}$ is rounded to either $\pm2^{-i}$ or $\pm2^{-i-1}$. The proof of
the second part of the lemma is the same as the first part. Here we obtain $x_1'$ and
$x_2'$ by the same procedure and follow the same argument to get $y_1$ and
$y_2$.
\end{proof}


\begin{lemma} \label{agp_log_lemma} 
Let $r\ge 2, x>1/2, z>0$ and $t$ be an integer such that $r^t \leq z/2$. 
Then, 
\begin{equation*}
\sum_{i=0}^{i=t} (r^i  \log(z/r^{i}))^x \leq  c(r)(r^t \log(z/r^t))^x
\end{equation*}
for a constant function $c(r)$ that depends only on $r$. Moreover, $c(2) < 9$.
\end{lemma}
\begin{proof}
Let $a_i:=(r^i  \log(z/r^{i}))^x\ \forall\ i=0,1,\ldots, t$. Let us consider the ratio of
consecutive terms $a_{i+1}/a_i$ for $i\in \{0,1,\ldots, t-1\}$.
\begin{align*}
\frac{a_{i+1}}{a_i} &= \left(\frac{r^{i+1}  \log(z/r^{i+1})}{r^i  \log(z/r^{i})}
\right)^x
\\ 
&= \left(r\left( 1 - \frac{\log(r)}{\log(z) - i\log(r)}\right)\right)^x \\
&\geq  \left(r\left( 1 - \frac{\log(r)}{1 + (t-i)\log(r)}\right)\right)^x &&
(r^t \leq z/2)
\end{align*}

If $i \leq t-2$, we get that $a_{i+1}/a_{i} \geq
r^x\left(\frac{1+\log(r)}{1 + 2\log(r)}\right)^x = \alpha(r)$. It is easy to
see that $\alpha(r) >\frac{2}{\sqrt{3}} > 1$ for $r \geq 2$. Also for $i=t-1$, we get that $a_{i+1}/a_{i}
\geq \left(r/(1+ \log(r))\right)^x \geq 1$. 
Now consider the sum $S_{-1}$ defined as 

\begin{align*}
&S_{-1} = a_0 + a_1 + \ldots + a_{t-1} \\
&\Rightarrow  \alpha(r) S_{-1} = \alpha(r)(a_0 + a_1 + \ldots + a_{t-1}) \\
&\Rightarrow  (\alpha(r) - 1)S_{-1} = -a_0 + (\alpha(r)a_0 - a_1) +
(\alpha(r)a_1 - a_2) \ldots + a_{t-1}\alpha(r) \\
&\Rightarrow  (\alpha(r) - 1)S_{-1} \leq a_{t-1}\alpha(r) && (a_{i+1} \geq
\alpha(r)a_i) \\
&\Rightarrow S_{-1} \leq a_{t-1}\left(\frac{\alpha(r)}{\alpha(r) - 1}\right)
\end{align*} 

Therefore $$\sum_{i \in [t]} a_i \leq S_{-1} + a_t \leq \left( 1 +
\left(\frac{\alpha(r)}{\alpha(r) - 1}\right) \right) a_t.$$

Setting $c(r) = \left( 1 + \left(\frac{\alpha(r)}{\alpha(r) - 1}\right)
\right)$ we get the identity. We observe that $\alpha(2)$ is greater than $\frac{2}{\sqrt{3}}$ which implies that $c(2) < 9$.
\end{proof}

\begin{fact}\label{fact:one}
For every $c_1 \geq 0$, there exists $c_2$ s.t. $\sqrt{\sqrt{x} \log \frac{1}{x}} \leq c_1 + c_2 x$ where $0 \leq x \leq 1$.
\end{fact}

\subsection{Spectral Graph Theory Basics}
Expander graphs are often seen as graphs which are close to random graphs.
This idea is quantified by the following well-known fact known as the Expander
Mixing Lemma which bounds the deviation between the number of edges between two
subsets and the expected number in a random graph.

\begin{theorem}[Expander-Mixing Lemma] \cite{Wigderson}\label{thm:EML} For a non-bipartite graph $G$,
\[ 
\left|E(S,T) - \frac{d|S||T|}{n}\right| \leq \lambda_G\sqrt{|S||T|}\ \ \forall\ S,T \subseteq V.
\]
\end{theorem}

We also have an analogue for bipartite graphs (by proceeding along the lines of the proof of the Expander Mixing Lemma). The following theorem states the general bound.
\begin{theorem}\label{thm:general-EML}
For a graph $G$, 
\[ 
E(S,T) \leq  2\frac{d|S||T|}{n} + \lambda_G \sqrt{|S| |T|}\ \ \forall\ S,T \subseteq V. 
\]
\end{theorem}

We need the following theorem showing that expanders have small diameter in order to show no-expansion of large abelian lifts. 
\begin{theorem} \cite{Chung89} \label{thm:diameter-and-eigenvalue}
The diameter of a d-regular graph $G$ with $n$ vertices is at most $\log(n)/\log(d/\lambda_G)$.
\end{theorem}

\section{Lifts}\label{subsec:lifts}

In this section we define lifts of graphs and state some of their properties. 


\begin{definition}[$(\Gamma,S,\cdot)$-lift]
Let $\Gamma$ be a group, $S$ be a set of size $k$ and $\cdot$ be a faithful group action of $\Gamma$ on $S$. A $(\Gamma,S,\cdot)$-lift of an $n$-vertex base graph $G=(V,E)$ is a graph $H=(V \times S, E')$ obtained as follows: it has $k\times n$ vertices, where each vertex $u$ of $G$ is replaced by $k$ vertices $\{u\}\times S$. For each edge $(u,v)$ of $G$, we choose an element $g_{u,v}\in \Gamma$ and replace that edge by a perfect matching between $\{u\}\times S$ and $\{v\}\times S$ that is given by the edges $(u_i,v_j)$ for which $g_{u,v}\cdot i = j$. We denote $|S|=k$ to be the order of the lift. 
\end{definition}

We note that if $S=\Gamma$ and the group action $\cdot$ is the left group operation itself, then $(\Gamma,S,\cdot)$-lifts are just $\Gamma$-lifts. 

\paragraph{Remark 5} (Group Elements as Permutations).\label{rem:premutations}
A faithful action of a group $\Gamma$ on a set $S$ induces an embedding from $\Gamma$ to $Sym(S)$, where $Sym(S)$ is the symmetric group of $S$ (group of all permutations of $S$). Thus, we can identify group elements with permutations of $|S|=k$ objects. Using this language, the set of edges of the lift $H$ can be rewritten as $E' = \{ (u_i,v_j) | (u,v) \in E, \pi_{u,v}(i)=j \} $, where $\pi_{u,v}$ is the permutation corresponding to the group element that we choose for edge $(u,v)$.\\

Besides $\Gamma$-lifts another interesting case of $(\Gamma,S,\cdot)$-lifts is when $\Gamma=Sym([k])$ (the symmetric group on $k$ elements), $S=[k]$ and the group action $\cdot:\Gamma\times S\rightarrow S$ is defined by $\sigma\cdot t=\sigma(t)$, i.e., the action of the permutation on the corresponding element. Such lifts are known as \emph{general lifts} or simply \emph{$k$-lifts}. Recall that shift $k$-lifts are $\Gamma$-lifts where the group $\Gamma$ is a cyclic group. We will use the term \emph{abelian lifts} to refer to $\Gamma$-lifts where the group $\Gamma$ is an abelian group.


Some initial easy observations can be made about the structure of any lift:
(i) the lifted graph is also regular with the same degree as the base graph and (ii) the eigenvalues of $A$ are also eigenvalues of $A_H$.
Therefore we call the $n$ eigenvalues of $A$ as the \emph{old} eigenvalues and
the $n(k-1)$ other eigenvalues of $A_H$ as the \emph{new} eigenvalues. We will denote by $\lambda_{new}$ the largest new eigenvalue of $H$ in magnitude, which we also refer to as the ``first'' new eigenvalue for simplicity.

\begin{definition}[Generalized Signing]
Given a base graph $G(V,E)$, a group $\Gamma$, a set $S$ and an action  $\cdot$ of $\Gamma$ on $S$ as in the above definition, we define a generalized signing of $G(V,E)$ as a function $s: E(G)\rightarrow \Gamma$. We use the convention that $s(u,v)=g$ then $s(v,u)= g^{-1}$. 
\end{definition}
We observe that there is a bijection between signings and $(\Gamma,S,\cdot)$-lifts.

\subsection{Spectrum of Lifts via Representation Theory}\label{sec:spectrum-of-lifts}
In this section, we characterize the spectrum of $\Gamma$-lifts as a union of the spectrum of certain matrices\footnote{On first read, some readers might want to skip ahead to Corollary \ref{lem:shiftspectrum}. 
Reading the details of this section is not essential for the purposes of understanding the results and the proofs in this work but it provides the main ideas for characterizing the spectrum of group--based lifts. Corollary \ref{lem:shiftspectrum} can also be shown by considering the eigenvectors directly.}.
We begin with some elementary facts on the representation theory of finite groups (see \cite{Artin, Serre}).  

\begin{definition}[Representation]
A representation of a finite group $\Gamma$ on a finite-dimensional vector space $\V$ is a homomorphism $\rho:\Gamma\rightarrow GL(\V)$, where $GL(\V)$ is the general linear group of $\V$. If the dimension of $\V$ is $\Delta$, then we define the dimension of $\rho$ to be $\Delta$.
\end{definition}

A trivial representation is one where $\mathcal{V}=\mathbb{C}$ and $\rho(g)=1$ for all $g\in \Gamma$. A \emph{permutation representation} is one where the matrices $\rho(g)$ correspond to permutation matrices. We next consider an interesting special case of permutation representations. 

\begin{definition}[Regular Representation]
For a group element $g\in \Gamma$, let $e_g$ be the $|\Gamma|$-dimensional indicator vector of $g$ and let $\C^{\Gamma}$ denote the vector space defined by the basis vectors $\{e_g\}_{g\in \Gamma}$. Let $P_g$ denote the permutation matrix associated with the left action of $g$ on $\Gamma$. Then $\rho(g)=P_g$ is a representation of $\Gamma$ on $\V=\C^{\Gamma}$.  This is known as the (left) regular representation of $\Gamma$ on $\C^{\Gamma}$.

\end{definition}

\begin{definition}[Irreducible Representation]
For a representation $\rho: \Gamma\rightarrow GL(\mathcal{V})$, a subspace $\mathcal{W} \subset \mathcal{V}$ is invariant under $\rho$ if $\rho(g)\mathcal{W}\subset \mathcal{W}$ for all $g\in \Gamma$. The representation $\rho$ is irreducible (hereafter called irrep) if it has no (proper) invariant subspace.
\end{definition}

A well-known theorem of Maschke shows that every permutation representation can be decomposed into a finite number of irreps. Our next theorem is a consequence of this result as applied to the regular representation.

\begin{theorem}[Decomposition into irreps for Regular Representation \cite{Serre}] \label{thm:decomposition-into-irreps}
Let $\rho$ be the regular representation of $\Gamma$ on $\C^{\Gamma}$. Then there exists a unitary matrix $U\in \C^{\Gamma\times \Gamma}$, an orthogonal decomposition $\C^{\Gamma}=\oplus \V_i$ and irreps $\rho_i:\Gamma\rightarrow GL(\V_i)$ such that $U\rho(g)U^{-1}=\oplus_i \rho_i(g)$ for every $g\in \Gamma$. 
Moreover, the trivial representation is always one of the irreps. 
\end{theorem}

We next state a few properties of the irreps arising in Theorem \ref{thm:decomposition-into-irreps} for abelian groups and cyclic groups. 

\begin{fact}\label{fact:irreps-for-cyclic-groups}
For abelian groups, the irreps in Theorem \ref{thm:decomposition-into-irreps} are one-dimensional. In particular, for a cyclic group $\Gamma=\{c,c^2,\ldots, c^k\}$, the irreps are given by $\rho_1,\ldots, \rho_k:\Gamma\rightarrow GL(\C)$, where $\rho_i(c^j)=\omega_i^j$, where $\omega_i$ is a primitive $k$-th root of unity. 
\end{fact}
We note that when $k=2$, the two roots of unity are $\omega_1=1$ and $\omega_2=-1$, and the only non-trivial irrep is $\rho_2$, where $\rho_2(0)=1, \rho_2(1)=-1$.

We now characterize the eigenvalues of $\Gamma$-lifts. We observe that the adjacency matrix of a $\Gamma$-lift is a $nk\times nk$ symmetric matrix, which has $n\times n$ blocks $B_{u,v}$, each of size $k\times k$; the block $B_{u,v}$ is the $k\times k$ zero matrix if $(u,v)$ is not an edge in $G$; for every edge $(u,v)$ of $G$, we have $B_{u,v} =P_{u,v}$, which is the permutation representation of the element $g =s(u,v) \in \Gamma$. The following theorem characterizes the spectrum of the lift in terms of the spectrum of certain smaller matrices. We note that even though $G$ is an undirected graph, for the purposes of the theorem, we view it as a directed graph where if $(u,v)\in E$ then $(v,u)\in E$. Recall that if $s(u,v)=g$, then $s(v,u)=g^{-1}$.

\begin{theorem}\label{thm:spectrum_lift}\cite{MS95}
For $g\in \Gamma$, let $G_g$ be the induced subgraph of $G$ consisting of (directed) edges $(u,v)\in E$ such that $s(u,v) =g$, and let $A_g$ be its adjacency matrix. The adjacency matrix of the lifted graph $H$ is equal to $A_H=\sum_{g \in \Gamma} A_g \otimes P_g =U \big(\oplus_i  \sum_{g\in \Gamma}A_g\otimes \rho_i(g) \big)U^{-1}$,  for some unitary matrix $U$. Here $\rho_i$ are the irreps of the regular (left) representation of $\Gamma$ given in Theorem \ref{thm:decomposition-into-irreps}. 
\end{theorem}

The above theorem shows that there is some basis given by the columns of the matrix $U$ such that $A_H$ is block-diagonal in that basis, with blocks $D_i=\sum_{g\in \Gamma} A_g\otimes \rho_i(g)$. In particular, the spectrum of $H$ is equal to the spectrum of the set of matrices $D_i$. We note that since for any group, $\rho_1$ is the trivial, one-dimensional representation, it follows that $D_1 =A_G$, the adjacency matrix of the original graph. This is consistent with the observation in Section \ref{subsec:lifts} that all the ``old'' eigenvalues of $G$ are also eigenvalues of $H$. 

We now specialize Theorem \ref{thm:spectrum_lift} to the case of cyclic groups to characterize the spectrum of shift $k$-lifts. For a shift $k$-lift of a graph $G=(V,E)$ with adjacency matrix $A$, which is given by the signing $(s(i,j)=g_{i,j})_{(i,j)\in E}$, define the following family of Hermitian matrices $A_s(\omega)$ parameterized by $\omega$ where $\omega$ is a primitive $k$-th root of unity. 

\[
	[A_s(\omega)]_{ij} = \begin{cases}
	0, & \mbox{if } A_{ij} = 0\\
	\omega^{g_{i,j}}, & \mbox{if } A_{ij} = 1
\end{cases}
\]

The following corollary regarding the spectrum of shift $k$-lifts follows from Theorem \ref{thm:spectrum_lift} and Fact \ref{fact:irreps-for-cyclic-groups}.

\begin{corollary}\label{lem:shiftspectrum}
Let $G=(V,E)$ be a graph and $H$ be a shift $k$-lift of $G$ with the corresponding signing of the edges $(s(i,j)=g_{i,j})_{(i,j)\in E}$, where $g_{i,j}\in C_k$. Then the set of eigenvalues of $H$ are given by 
\[
\bigcup_{\omega:\ \omega \text{ is a primitive $k$-th root of unity}} \text{eigenvalues}\left(A_s(\omega)\right).
\]
\end{corollary}

The above simplifies significantly for  2-lifts as noted in the next corollary.
\begin{corollary}\label{lem:twospectrum}
When $k=2$, the set of eigenvalues of a 2-lift $H$ is given by the eigenvalues of $A$ and the eigenvalues of $A_s$, where $A_s$ is the signed adjacency matrix corresponding to the signing $s$, with entries from $\{0,1,-1\}$.
\end{corollary}


\section{No-expansion of Abelian Lifts}\label{sec:abelian-lifts}

In this section we show that it is impossible to find (even slightly) expanding graphs using lifts in large abelian groups $\Gamma$. By Theorem \ref{thm:diameter-and-eigenvalue}, we know that if a graph is an expander, then it has small diameter. We show that if the size of the (abelian) group $\Gamma$ is large, then \emph{all} $\Gamma$-lifts of any base graph have large diameter, and hence they cannot be expanders. We prove Theorem \ref{thm:abelian_lift_not_exists} that is restated here.

\thmAbelian*

\begin{proof}
We prove the contrapositive. Let $\Gamma$ be an abelian group of order $k$ and $G=(V,E)$ be a base graph on $n$-vertices that is $d$-regular. Let $e_1,\ldots, e_{nd/2}$ be an arbitrarily chosen ordering of the edges $E$. Let $H$ be a lift graph obtained using a $\Gamma$-lift. Recall that the signing of the edges of the base graph correspond to group elements, which in turn correspond to permutations of $k$ elements. Let these signing of the edges be $(\sigma_e)_{e \in E(G)}$. Let us define a layer $L_i$ of $H$ to be the set of vertices $\{v_i:v\in V\}$. We note that $H$ has $k$ layers. 

Let us fix an arbitrary vertex $v$ in $G$. Let $\Delta$ denote the diameter of $H$. This implies that 
for every $j=2,\ldots,k$ there exists a path of length at most $\Delta$ in $H$ from $v_1$ to a vertex in $L_j$. A layer $j$ is reachable within distance $\Delta$ in $H$ iff
there exists a walk $e_1,e_2,\ldots, e_t$ from $v$ of length $t\le \Delta$ in $G$ such that $\sigma_{e_t}\sigma_{e_{t-1}}\ldots\sigma_{e_2}\sigma_{e_1}(1)=j$. Thus the set of layers reachable within distance $\Delta$ in $H$ is contained in the set $S=\{\sigma_{e_t}\ldots\sigma_{e_1}(1): e_1,\ldots, e_t\text{ is a walk from $v$ in $G$ of length $t\le \Delta$}\}$. Since the group $\Gamma$ is abelian, 
$S\subseteq \{\sigma_{e_1}^{a_1} \sigma_{e_2}^{a_2} \dots \sigma_{e_{nd/2}}^{a_{nd/2}}(1) \mid \sum_{i=1}^{nd/2} |a_i | \leq \Delta\}=:T$. Since $H$ has $k$ layers, the cardinality of $S$ is at least $k$.

The number of integral $a_i$'s satisfying $\sum_{i=1}^{nd/2} |a_i | \leq \Delta$ is at most ${(nd/2)+\Delta \choose (nd/2)}\cdot 2^{(nd/2)}$. Therefore,
\begin{align*}
k\le |T|
\le \binom{\frac{nd}{2}+\Delta}{\frac{nd}{2}} 2^{\frac{nd}{2}}
&\le \left(2e\left(1+\frac{2\Delta}{nd}\right)\right)^{\frac{nd}{2}}
\le (2e)^{\frac{nd}{2}} e^{\Delta}.
\end{align*}

Since $H$ has $nk$ vertices, using Theorem \ref{thm:diameter-and-eigenvalue}, we have $\Delta\le (\log nk)/\log(d/\lambda(H))$. Thus, if $\lambda(H) \le \epsilon d$, then $\Delta\le (\log nk)/\log(1/\epsilon)$ and consequently, 
\begin{align*}
k
&\le (2e)^{\frac{nd}{2}} e^{\frac{\log{nk}}{\log{\frac{1}{\epsilon}}}}.
\end{align*}

Rearranging the terms, we obtain that 
\[
k
\le (2e)^{\frac{nd}{2\left(1-\frac{1}{\log{\frac{1}{\epsilon}}}\right)}} exp\left(\frac{\log{n}}{\left(\log{\frac{1}{\epsilon}}\right)\left(1-\frac{1}{\log{\frac{1}{\epsilon}}}\right)}\right)
\le exp\left(\frac{nd\log\frac{1}{\epsilon}+\log{n}}{\log{\frac{1}{e\epsilon}}}\right).
\]



\end{proof}
	
\section{Expansion of Random $2$-lifts: Overview}\label{sec:proofs_short}


In this section, we illustrate the main techniques involved in proving Theorem \ref{thm:main2}
by stating and proving 
a slightly weaker version, namely Theorem \ref{main_theorem_easy}. It focuses only on $2$-lifts akin to Corollary \ref{thm:main1} and is weaker in comparison to the bound in Corollary \ref{thm:main1} by a multiplicative factor of four. The proof of this weaker result captures the main ideas involved in the proof of Theorem 
\ref{thm:main2}. 

\begin{theorem}\label{main_theorem_easy}
Let $G$ be a $d$-regular $n$-vertex graph with non-trivial eigenvalues at most $\lambda$
in absolute value where $\lambda\ge \sqrt{d}$, $2 \le d \leq \sqrt{n/(3\ln n)}$, and $H$ be a uniformly random $2$-lift of $G$. Let
$\lambda_{new}$ be the largest new eigenvalue of $H$ in magnitude. Then, 
$$\lambda_{new} \le 4\lambda + 
4\cdot 10^{13}
\max\left(\sqrt{\lambda \log d},\sqrt{d}\right)$$ 
with probability at least $1-e^{-n/d^2}$.
\end{theorem}
To prove this theorem, we require the following concentration inequality. It is derived from Hoeffding's inequality by taking a suitable union bound. 
We present the complete proof in Section \ref{sec:proof_lemma}.

\begin{restatable}{lemma}{lemMain}\label{main_lemma}
Let $G$ be a $d$-regular graph with non-trivial eigenvalues at most $\lambda$ in absolute value where $\lambda\ge \sqrt{d}$, $2 \le d \leq\sqrt{n/(3\ln n)}$. Let $H$ be a uniformly random $2$-lift of $G$, with corresponding signed adjacency matrix $A_s$. The following statements hold with probability at least $1- e^{-n/d^2}$ over the choice of the random signing:

\begin{enumerate}
\item For all $u_1,\ldots,u_r \in \{0,\pm 1\}^n$, and $v_1,\ldots,v_{\ell} \in \{0,\pm 1\}^n$ satisfying
\begin{enumerate}
\item[(I)] $S(u_i)\cap S(u_j)=\emptyset$ for every $i,j\in [r]$ and $S(v_i)\cap S(v_j)=\emptyset$ for every $i,j\in [\ell]$, and
\item[(II)] Either $|S(u_i)|> n/d^2$ for every $i\in [r]$ with non-zero $u_i$, or $|S(v_i)|> n/d^2$ for every $i\in [\ell]$ with non-zero $v_i$,
\end{enumerate}
we have
\begin{equation*}
\left|\sum \limits_{i \leq j} (2^{-i}u_i^T) A_s (2^{-j}v_j)\right| 
\le 377\max(\sqrt{\lambda \log d},\sqrt{d})
\sum_{i=1}^r |S(u_i)| 2^{-2i} +
\left(\frac{\lambda}{5} + 10^{12}\sqrt{d}\right)\sum_{j=1}^{\ell}|S(v_j)| 2^{-2j}.
\end{equation*}
\item For all $u_1,\ldots, u_r\in \{0,\pm 1\}^n$, and $v_1,\ldots,v_{\ell} \in \{0,\pm 1\}^n$ satisfying (I), (II) and
\begin{enumerate}
\item[(III)] $|S(u_i)|>|S(v_j)|$ for every $i\in[r],j\in[\ell]$ with non-zero $u_i$,
\end{enumerate}
we have
\begin{equation*}
\left|\sum\limits_{i \leq j} (2^{-i}u_i^T) A_s (2^{-j}v_j)\right| = 31\max\left(\sqrt{\lambda \log d},\sqrt{d}\right)
\left(\sum_{i=1}^r |S(u_i)| 2^{-2i} + \sum_{j=1}^{\ell}|S(v_j)| 2^{-2i}\right).
\end{equation*}
\end{enumerate}
\end{restatable}

We will now prove Theorem \ref{main_theorem_easy} using the lemma above.

\begin{proof}[Proof of Theorem \ref{main_theorem_easy}]
By Corollary \ref{lem:twospectrum}, the largest new eigenvalue (in absolute value) of the lift is $\LN=\max_{x\in \R^n}|x^TA_sx/x^Tx|$. To prove an upper bound on $\LN$, we will bound  $|x^TA_sx/x^Tx|$ for all $x$ with high probability. In particular, assuming that the concentration inequalities given by Lemma \ref{main_lemma} holds, we will show that 
\[\left|x^T A_s x \right| \leq  4\left(\lambda + 10^{13}\sqrt{d}\right)\|x\|^2.\]


By re-scaling we may assume that the maximum entry of $x$ is less than $1/2$ in absolute value. By Lemma \ref{DiscretizationLemma}, there exists a vector $y \in \{\pm 1/2, \pm 1/4, \ldots\}^n$ such that $|x^TA_sx| \leq |y^TA_sy|$ and $\|y\|^2\leq 4\|x\|^2$. We will prove a bound on $|y^TA_sy|$ for every $y\in\{\pm 1/2,\pm 1/4,\ldots\}^n$, which in turn will imply the desired bound on $|x^T A_s x|$. Let us consider the diadic decomposition of $y = \sum_{i=1}^{\infty} 2^{-i} u_i$ obtained as follows: a coordinate of $u_i$ is $1$ if the corresponding coordinate of $y$ is $2^{-i}$, it is $-1$ if the corresponding coordinate in $y$ is $-2^{-i}$, and is zero otherwise. We note that $S(u_i)\cap S(u_j)=\emptyset$ for every pair $i,j\in \mathbb{N}$.

Next, we partition the set of vectors $u_i$'s based on their support sizes. Let $M := \{ i\in\mathbb{N} : |S(u_i)| \leq n/d^2 \}$ and $L := \{ i\in\mathbb{N} : |S(u_i)| > n/d^2\}$ ($M$ and $L$ for mini and large supports respectively). Correspondingly, define $y_M := \sum_{i\in M} 2^{-i} u_i$ and $y_L = \sum_{i\in L} 2^{-i} u_i$. We note that $y = y_M + y_L$, $\|y\|^2 = \|y_M\|^2 + \|y_L\|^2 = \sum_{i\in \N} |S(u_i)| 2^{-2i}$, and

\begin{equation*}
|y^T A_s y| \leq |y_M^T A_s y_M| + 2 |y_M^T A_s y_L| + |y_L^T A_s y_L|.
\end{equation*}
We next bound each term in the following three claims.

\begin{claim}\label{claim:helper1}
\[ |y_M^T A_s y_M| \le \left(\lambda +\frac{8}{d}\right)\|y_M\|^2. \]
\end{claim}
\begin{proof}
Let $y_M'$ be a vector obtained from $y_M$ by taking the absolute values of each entry. 
Then $\|y_M\|^2 = \|y_M'\|^2$ and $|y_M^T A_s y_M| \leq y_M'^T A y_M'$. Let $J = v v^T$ and $J' = v' v'^T$ where $v$ is all 1 vector and $v'$ is defined as follows: $v'_i = 1$ for $1\leq i \le n/2$ and $v'_i = -1$ for $n/2 + 1 \le i \le n$. For non-bipartite graph $G$, we have 
\begin{equation*}
y_M'^T A y_M' =
y_M'^T \left(A-\frac{d}{n}J\right)y_M' + y_M'^T \left(\frac{d}{n}J\right)y_M' \leq 
\lambda \|y_M'\|^2 + y_M'^T \left(\frac{d}{n}J\right)y_M'.
\end{equation*}

Above, we have used the fact that $A- \frac{d}{n} J$ has the same set of eigenvalues as $A$ except for the first eigenvalue which was $d$ for the matrix $A$ and is now zero.  Similarly, for bipartite graphs, we have
\begin{equation*}
y_M'^T A y_M' =
y_M'^T \left(A-\frac{d}{n}J + \frac{d}{n} J'\right)y_M' + y_M'^T \left(\frac{d}{n}J\right)y_M' - y_M'^T \left(\frac{d}{n}J'\right)y_M' \leq 
\lambda \|y_M'\|^2 + y_M'^T \left(\frac{d}{n}J\right)y_M' - y_M'^T \left(\frac{d}{n}J'\right)y_M'.
\end{equation*}
Above, we have used the fact that $A-\frac{d}{n}J + \frac{d}{n} J'$ has the same set of eigenvalues as $A$ except the largest two eigenvalues (in absolute value) of $A$ which were $d$ and are now zero. It remains to bound $|y_M'^T \left(\frac{d}{n}J\right) y_M'|$ and $| y_M'^T \left(\frac{d}{n}J'\right) y_M'|$.
Consider the diadic decomposition of $y_M' = \sum_{i \in M} 2^{-i}u_i'$, where the coordinates of $u_i'$ are the absolute values of the coordinates of $u_i$. 
\begin{align*}
\left|y_M'^T \left(\frac{d}{n}J\right) y_M'\right|, \left|y_M'^T \left(\frac{d}{n}J'\right) y_M'\right| 
&\leq 2\sum_{i\in M}\sum_{j\in M:j \geq i} \frac{d}{n}2^{-i}|S(u_i)|2^{-j}|S(u_j)| \\
&\leq 2\sum_{i\in M} \frac 1d 2^{-2i}|S(u_i)| \sum\limits_{j\in M:j \geq i} 2^{i-j} \quad \quad \text{(since $|S(u_j)|\le n/d^2\ \forall\ j\in M$)}\\
&\leq \frac{4}{d}\|y_M'\|^2.
\end{align*}

\end{proof}

\begin{claim}\label{claim:helper2}
\[
|y_L^T A_s y_L| \le  \left(\frac{2\lambda}{5} + (3\cdot 10^{12})\max\left(\sqrt{\lambda \log d},\sqrt{d}\right)\right)\|y_L\|^2.
\]
\end{claim}
\begin{proof}
By triangle inequality, 
\begin{align*}
|y_L^T A_s y_L| 
&=  \left|\sum_{i,j \in L} (2^{-i} u_i^T) A_s (2^{-j}u_j)\right|\\
& \leq \left|\sum \limits_{i,j\in L:i\le j} (2^{-i}u_i) A_s (2^{-j}u_j)\right| + \left|\sum \limits_{i,j\in L:i > j} (2^{-i}u_i) A_s (2^{-j}u_j)\right|.
\end{align*}

We bound each term using the first part of Lemma \ref{main_lemma}. For both terms, our choice is $r\leftarrow \max\{i\in L\}$, $\ell=r$, $u_i\leftarrow u_i$ if $i\in L$ and $u_i\leftarrow \overline{0}$ if $i\not\in L$, $v_i=u_i$ for every $i\in [r]$, where $\overline{0}$ is the all-zeroes vector. We note that the conditions (I) and (II) of Lemma \ref{main_lemma} are satisfied by this choice since every pair $S(u_i), S(u_j)$ is mutually disjoint and $|S(u_i)| > \frac{n}{d^2}$ for all $i \in L$. Consequently, 

\begin{align*}
|y_L^T A_s y_L| 
&\le 754\max\left(\sqrt{\lambda \log d},\sqrt{d}\right)
\sum\limits_{i \in L} |S(u_i)| 2^{-2i} + \left(\frac{\lambda}{5} + 2\cdot 10^{12}
\sqrt{d}\right)\sum\limits_{j \in L} |S(u_j)| 2^{-2j} \\
&\le \left(\frac{2\lambda}{5} + (2\cdot 10^{12} + 754)\max\left(\sqrt{\lambda \log d},\sqrt{d}\right)\right) \|y_L\|^2.
\end{align*}
\end{proof}

\begin{claim}\label{claim:helper3}
\[
|y_M^T A_s y_L| \le 408\max\left(\sqrt{\lambda \log d},\sqrt{d}\right) \|y_M\|^2 +  \left(\frac{\lambda}{5} +(2\cdot 10^{12})\max\left(\sqrt{\lambda \log d},\sqrt{d}\right)\right) \|y_L\|^2.
\]
\end{claim}
\begin{proof}
By triangle inequality, 
\begin{align*}
|y_M^T A_s y_L| 
&= \left|\sum_{i\in M,j \in L} (2^{-i} u_i^T) A_s (2^{-j}u_j)\right| \\
&\leq \left|\sum \limits_{i\in M, j\in L:i \leq j} (2^{-i}u_i) A_s
(2^{-j}u_j)\right| + \left|\sum \limits_{i\in M, j\in L:i > j} (2^{-i}u_i) A_s 
(2^{-j}u_j)\right|.
\end{align*}

We bound the first and second terms by the first and second parts of Lemma \ref{main_lemma} respectively. Let $\overline{0}$ be the all-zeroes vector.
For the first term, our choice is $r\leftarrow \max\{i\in M\}$, $\ell\leftarrow \max\{i\in L\}$, $u_i\leftarrow u_i$ if $i\in M$ and $u_i\leftarrow \overline{0}$ if $i\not\in M$, and
$v_i\leftarrow u_i$ if $i\in L$ and $v_i\leftarrow \overline{0}$ if $i\not\in L$.
For the second term, our choice is $r\leftarrow \max\{i\in L\}$, $\ell\leftarrow \max\{i\in M\}$, $u_i\leftarrow u_i$ if $i\in L$ and $u_i\leftarrow \overline{0}$ if $i\not\in L$, and
$v_i\leftarrow u_i$ if $i\in M$ and $v_i\leftarrow \overline{0}$ if $i\not\in M$. The conditions (I), (II) and (III) of Lemma \ref{main_lemma} are satisfied for the respective choices since every pair $S(u_i), S(u_j)$ is mutually disjoint, $|S(u_i)| > \frac{n}{d^2}$ for all $i \in L$ and $|S(u_i)|> n/d^2\ge |S(u_j)|$ for every $i\in L, j\in M$. Consequently,

\begin{align*}
|y_M^T A_s y_L| 
&\le 377\max\left(\sqrt{\lambda \log d},\sqrt{d}\right)
\sum\limits_{i \in M} |S(u_i)| 2^{-2i} + \left(\frac{\lambda}{5} + 10^{12}\sqrt{d}\right)\sum\limits_{j \in L} |S(u_j)| 2^{-2j} \\
& \quad \quad + 31\max\left(\sqrt{\lambda\log d},\sqrt{d}\right) \left(\sum\limits_{j \in L} |S(u_j)| 2^{-2j} + \sum\limits_{j \in M} |S(u_j)|2^{-2j}\right)\\
&\le 408\max\left(\sqrt{\lambda \log d},\sqrt{d}\right) \|y_M\|^2 +  \left(\frac{\lambda}{5} +(10^{12}+ 31)\max\left(\sqrt{\lambda \log d},\sqrt{d}\right)\right) \|y_L\|^2.
\end{align*}
\end{proof}


From the above three claims, we have 
\begin{align*}
|y^T A_s y|
&\le \left(\lambda + 817\max\left(\sqrt{\lambda \log d},\sqrt{d}\right)\right)\|y_M\|^2 + \left(\frac{4\lambda}{5} + (7\cdot 10^{12} )\max\left(\sqrt{\lambda \log d},\sqrt{d}\right)\right) \|y_L\|^2\\ 
&\le \lambda + 8\cdot 10^{12}\max\left(\sqrt{\lambda \log d},\sqrt{d}\right)\|y\|^2.
\end{align*}
Therefore, we have
\begin{eqnarray*}
|x^T A_s x| &\leq& |y^T A_s y| \\
&\le& \left(\lambda +
8\cdot 10^{12}\max\left(\sqrt{\lambda \log d},\sqrt{d}\right)\right)\|y\|^2 \\
&\le & 4\left(\lambda +
8\cdot 10^{12}\max\left(\sqrt{\lambda \log d},\sqrt{d}\right)\right)\|x\|^2.
\end{eqnarray*}
\end{proof}

We note that in the above proof, the multiplicative factor of 4 is a by-product of the discretization of $x$. This can be avoided if we do not discretize $x$ straightaway, but instead ``push'' the discretization a little deeper into the proof. Indeed, we can see that the proof of Claim \ref{claim:helper1} where we bound $|y_M^T (A-\frac{d}{n}J) y_M|$ by $\lambda \|y_M\|^2$ does not require $y_M$ to be a discretized vector. This is how we are able to prevent the multiplicative factor loss to obtain Theorem \ref{thm:main2}.

\section{Concentration Inequality}
\label{sec:proof_lemma}
In this section, we prove Lemma \ref{main_lemma}. 
In order to prove Lemma \ref{main_lemma} we need to upper bound $$\left|\sum_{i \leq j}2^{-i-j} u_i^T A_s v_j\right|$$ for all sets of vectors $\{u_1,\ldots, u_r\}$, $\{v_1,\ldots, v_{\ell}\}$ satisfying the assumptions of the lemma over random choices of $A_s$. 
A natural approach is to 
use the triangle inequality and upper bound each term $|u_i^T A_s v_j|$ separately for each $i,j$. We note that $u_i^T A_s v_j$ is a sum of $|E(S(u_i),S(v_j))|$ iid random variables with mean zero (one for each edge between $S(u_i)$ and $S(v_j)$). By the expander mixing lemma (Theorem \ref{thm:general-EML}), we may upper bound the size of $E(S(u_i),S(v_j))$ by $2d |S(u_i)||S(v_j)|/n + \lambda \sqrt{|S(u_i)| |S(v_j)|}$. Depending on which of these two terms in the RHS dominates, we have two cases. For each case, we use a different concentration bound (Lemma \ref{prob_lemma_2} and Corollary \ref{alternative_prob_lemma_corollary}). We begin with the needed concentration bounds.


\subsection{Concentration Bounds}

\begin{lemma}
\label{prob_lemma_2}
Let $G$ be a $d$-regular, $n$-vertex graph with non-trivial eigenvalues at most $\lambda$ in absolute value where $\lambda \ge \sqrt{d}$, $2\leq d \leq \left(2n/3\ln n\right)^2$. Let $H$ be a uniformly random $2$-lift of $G$, with corresponding signed adjacency matrix $A_s$. 
The following property holds with probability atleast $1-e^{-(n\log{d})/\sqrt{d}}$ (over the random choice of signings):

For every $r\in\{0,1,\ldots,(1/2)\log{d}\}$, every $a,b_{0},b_{1},\dots,b_r \in \{0,\pm1\}^n$ satisfying
\begin{enumerate}
\item[(i)] $S(b_i)\cap S(b_j)=\emptyset\ \forall\ i,j\in [r],i\neq j $,
\item[(ii)] $|S(a)| \geq 2^{2i}|S(b_i)|\ \forall\ i\in [r]$, and 
\item[(iii)]$\frac{d}{\lambda} \sqrt{|S(b_i)||S(a)|} \geq n\ \forall\ i\in [r]$ with non-zero $b_i$,
\end{enumerate}
we have 
\begin{equation*} \label{eq:prob_condition_2}
\left|a^T A_s \left(\sum_{i=0}^r2^ib_i\right)\right| \leq 
14 \sqrt{\frac{d}{n} |S(a)|^2\left(\sum_{i=0}^r |S(b_i)|2^{2i}\right)\log\left(\frac{2n}{|S(a)|}\right)}.
\end{equation*}
\end{lemma}
\begin{proof}
For notational convenience, let $b=\sum_{i=0}^r 2^ib_i$. 
Fix $a,b_1,b_2,\ldots, b_r \in \{0,\pm 1\}^n$. Then $a^TA_s b$ is a sum of
independent random variables with mean $0$ one for each edge between $S(a)$ and $S(b_i)$. This is because the
intersection between the support of any two vectors $b_i$ and $b_j$ is empty. The
sum of squares of the difference between the maximum and the minimum values of these variables is at most $\sum_{i=1}^r
4E(S(b_i),S(a))2^{2i}$. For vectors $a,b_1,\ldots, b_r$ satisfying (i) and (ii), 
by the Expander Mixing Lemma, we have $E(S(b_i),S(a)) \leq 3\frac{d|S(b_i)||S(a)|}{n}$. We note that this inequality holds even if $b_i$ is a zero vector.

By Theorem \ref{thm:hoeffding_inequality}, 
\[ Pr \left( |a^TA_s b| > 14 \sqrt{\frac{d}{n} |S(a)|^2\left(\sum_i |S(b_i)|2^{2i}\right)
\log\left(\frac{2n}{|S(a)|}\right)} \right)  \leq 2exp \left( -\frac{98|S(a)|}{3}\log\left(\frac{2n}{|S(a)|}\right) \right)
\]

Now fixing the values of the support sizes $\alpha=|S(a)|,\beta_i=|S(b_i)|$, the number of possible choices for $a$ is at most $\binom{n}{\alpha}*2^{\alpha} \leq
exp \left( 3\alpha\log(\frac{2n}{\alpha}) \right)$. Similarly the number of
possible choices for each $b_i$ is atmost $\exp \left(
3\beta_i\log(\frac{2n}{\beta_i}) \right)$.
Therefore the total number of choices for $b$ is at most $
exp \left( \sum_{i=1}^r 3\beta_i\log(\frac{2n}{\beta_i}) \right) $. Since
each $\alpha,\beta_i\leq n$, we can replace each $\beta_i$ by its upper bound
$\alpha2^{-2i}$. Hence, using Lemma \ref{agp_log_lemma},

\begin{eqnarray*}
\exp \left( \sum_{i=1}^r 3\beta_i\log\left(\frac{2n}{\beta_i}\right) \right) &\leq&
exp \left( 3 \sum_{i=1}^r \alpha{2^{-2i}}\log(\frac{2n}{\alpha2^{-2i}}) \right) 
\leq exp \left( 27\alpha\log\left(\frac{2n}{\alpha}\right) \right). 
\end{eqnarray*}


Therefore, the total number of choices of $a,b_1, \ldots, b_r$ of sizes $\alpha,\beta_1,
\ldots \beta_r$ respectively is at most $$\exp \left( 30\alpha\log\left(\frac{2n}{\alpha}\right)
\right).$$
By taking a union bound over the choices of vectors with the fixed support sizes, the probability of the existence of a set of vectors $a,b_1,\ldots, b_r$ with sizes $\alpha,\beta_1,\ldots, \beta_r$ respectively  and satisfying (i) and (ii) is bounded by 

\[ 2exp \left( -\frac{8\alpha}{3}\log\left(\frac{2n}{\alpha}\right) \right) \leq 2exp \left(
-\frac{8n}{3\sqrt{d}}\log{(2\sqrt{d})} \right). \]
Above, we have used that $\alpha\ge n/
\sqrt{d}$ which follows since $\alpha=|S(a)|\ge n\lambda/d\ge n/\sqrt{d}$ by $(ii)$ and $(iii)$. Next, let us bound the number of choices for the support sizes of the vectors $a,b_1,\ldots, b_r$. 
The number of choices for the support sizes is at most $n^{2+(1/2) \log d}$. 
Therefore taking the union bound over the choice of the support sizes, we get that
the total probability is at most 
\[ 2exp \left( (2+(1/2) \log d)\ln{n} \right) exp \left(
-\frac{8n}{3\sqrt{d}}\log(2\sqrt{d}) \right) \leq
exp \left( -\frac{n\log{d}}{\sqrt{d}} \right). \]

\end{proof}

\begin{lemma}
\label{alternative_prob_lemma_1}
Let $G$ be a $d$-regular, $n$-vertex graph with non-trivial eigenvalues at most $\lambda$ in absolute value, where $2\le d\le \sqrt{n/3 \ln n}$, and $H$ be a uniformly random $2$-lift of $G$, with corresponding signed adjacency matrix $A_s$. The
following property holds with probability at least $1-e^{-3n/d^2}$ (over the random choice of signings):

For every $a,b \in \{0,\pm 1\}^n, q,w\in\{1,\dots,n\} $ satisfying 
\begin{enumerate}
\item[(i)] $|S(a)| \leq q$, $|S(b)| \leq w$, $S(b) \subset N_G(S(a))$,
\item[(ii)] $q\leq w\leq d q$, 
\item[(iii)] $w >\frac{n}{d^2}$, and 
\item[(iv)] $\frac{d}{\lambda} \sqrt{q  w} < n$, 
\end{enumerate}
we have 
\begin{equation} \label{eq:alternative_prob_condition_1}
|a^T A_s b| \leq 10 \sqrt{\lambda \sqrt{q } w^{3/2} \log\left(\frac{2
d  q}{w}\right)}.
\end{equation}
Here, $N_G(S(a))$ denotes the set of neighbors of $S(a)$ formally defined as $\{v \mid \exists u \in S(a)\text{ with } (u,v) \in E\}$.
\end{lemma}
\begin{proof}
For a pair of vectors $a,b\in \{0,\pm 1\}^n$ and $q,w \in\{1,\dots,n\}$, let $Bad(a,b,q,w)$ denote the event that inequality (\ref{eq:alternative_prob_condition_1}) is violated. We need to upper bound the probability that there exists $(a,b,q,w)$ satisfying $(i)$, $(ii)$, $(iii)$ and $(iv)$ such that $Bad(a,b,q,w)$ happens. We note that the sum $a^T A_s b$  over random choices of $A_s$ is a sum of
independent random variables chosen from $\{\pm 2, \pm 1\}$, all of which have mean $0$. The number of such random variables being summed is at most $E(S(a),S(b))$, i.e. the number of
edges between $S(a)$ and $S(b)$.

Therefore for a fixed $a,b,q,w$ by applying the Hoeffding inequality (Theorem \ref{thm:hoeffding_inequality}), we get that
\[
 P(Bad(a,b,q,w)) 
\leq
2exp \left( -\frac{50\lambda \sqrt{q} w^{3/2} \log \left( \frac{2
d q}{w} \right) }{E(S(a),S(b))} \right).
\]

Now using (iv) and the expander mixing lemma (Theorem \ref{thm:general-EML}), we have 

\[ E(S(a),S(b)) \leq 2 d  |S(a)| |S(b)|/n + \lambda \sqrt{|S(a)| |S(b)|} \leq 2 d q w/n + \lambda \sqrt{q w} \leq 3\lambda\sqrt{q w}.\] 

Substituting this in the previous expression, we obtain 

\[ P(Bad(a,b,q,w)) \leq 2 exp\left(-(50/3) w \log \left( \frac{2d  q}{w} \right) \right). \]

We will use the union bound now. For this purpose, we will first fix $q,w$ and the size of the support of $a$ and $b$. We take a union bound over all possible choices of $a,b$ of that fixed size, and then take a union bound
over all choices of the support sizes. For fixed support sizes $\alpha=|S(a)|,\beta=|S(b)|$, we observe that the total number of choices for the support sets for $a$ are $\binom{n}{\alpha}$. Now, since $S(b)$ is a subset of $N_G(S(a))$, the number of choices of $S(b)$ is bounded by
$\binom{d \alpha }{\beta}$. Also, since each entry in $a,b$ is $0$ or $\pm 1$  the total
number of choices for $a$ and $b$ is at most 
 
\[
 \binom{n}{\alpha}2^{\alpha}\binom{d\alpha}{\beta}2^{\beta} 
\leq
exp \left(3\alpha\log\left( \frac{2n}{\alpha} \right) \right)
 exp \left(3\beta\log\left(\frac{2d\alpha}{\beta}\right) \right)
\]

We will first show upper bounds on each of these terms. Since
$w\geq \frac{n}{d^2}$, by (ii), we have $q \geq\frac{n}{d^3}$. Also, $\alpha = |S(a)| \leq q, \beta= |S(b)| \leq w$. Therefore, 

\begin{eqnarray*}
exp \left(3\alpha\log\left( \frac{2n}{\alpha} \right)\right)
& \leq & exp \left(3q\log\left( \frac{2n}{q} \right) \right)\\
&\leq& exp\left( 9q\log(2d) \right) \\
&=& exp \left(9\frac{\frac{q}{w}\log(2d)}{\log
\left(2d\frac{q}{w}\right)}\cdot w\log \left(2d\frac{q}{w}
\right) \right) \\
&\leq& exp\left(9w\log\left(2d\frac{q}{w}\right)\right).
\end{eqnarray*}

The last line follows from the fact that $x\log(d)/\log(2dx)$ is
bounded by $1$ for $x \in [1/d,1]$ and that $\frac{q}{w} \in [
1/d,1 ]$. Further, 
\begin{align*}
exp \left(3\beta\log\left(\frac{2d\alpha}{\beta}\right) \right) 
& \leq  exp\left( 3\beta \log\left( \frac{2dq}{\beta}\right)\right) 
 \leq exp \left(3w\log\left(\frac{2dq}{w} \right)\right). 
\end{align*}
The last inequality follows by the fact that $x \log \frac{2c}{x}$ is an increasing function if $x<c$. Therefore, by union bound we get that the probability of a bad event for fixed $q,w$ and 
support sizes $\alpha=|S(a)|,\beta=|S(b)|$ is at most 

\begin{equation*}
2exp\left(-(14/3)w\log{\frac{4dq}{w}}\right)
\le 2exp \left(-\frac{14n}{3d^2}\log{\frac{4dq}{w}}\right)
\le 2exp \left( - \frac{14n}{3d^2} \log{2} \right).
\end{equation*}

Now the number of choices of the supports is at most $n^2$, number of choices for $q,w$ is at most $n^2$ and therefore, 

\begin{equation*} 
P(\exists (a,b,q,w) \text{ satisfying $(i)$, $(ii)$, $(iii)$, and $(iv)$:} Bad(a,b,q,w))
\le 2n^4exp \left( - \frac{14n}{3d^2} \log{2} \right)
\le exp\left(-\frac{3n}{d^2}\right).
\end{equation*}
\end{proof}

\begin{corollary}
\label{alternative_prob_lemma_corollary}
Let $G$ be a $d$-regular, $n$-vertex graph with non-trivial eigenvalues at most $\lambda$ in absolute value, where $2\le d\le \sqrt{\frac{n}{3 \ln n}}$, and $H$ be a uniformly random $2$-lift of $G$, with corresponding signed adjacency matrix $A_s$. The
following property holds with probability at least $1-e^{-3n/d^2}$ (over the random choice of signings):

For every $a,b \in \{0,\pm 1\}^n$ satisfying 
\begin{enumerate}
\item[(i)] $|S(a)| \leq |S(b)| \leq d|S(a)|$, 
\item[(ii)] $|S(b)| >\frac{n}{d^2}$, and 
\item[(iii)] $\frac{d}{\lambda} \sqrt{|S(a)||S(b)|} < n$, 
\end{enumerate}
we have 
\begin{equation} \label{eq:prob_condition_1}
|a^T A_s b| \leq 10 \sqrt{\lambda \sqrt{|S(a)| |S(b)|} |S(b)| \log\left(\frac{2
d|S(a)|}{|S(b)|}\right)}.
\end{equation}
\end{corollary}
\begin{proof}

For every $a,b$, we apply the bound from Lemma \ref{alternative_prob_lemma_1} on $|a^T A_s b'|$ with $q = |S(a)|, w = |S(b)|$ where $b'$ is the same as $b$ restricted to the coordinates in $S(b) \cap N_G(S(a))$. We observe that $|a^T A_s b| = |a^T A_s b'|$ and hence the corollary.

\end{proof}




\subsection{Proof of {Lemma} \ref{main_lemma}}
Next, we use Corollary \ref{alternative_prob_lemma_corollary} and Lemma \ref{prob_lemma_2} to prove Lemma \ref{main_lemma}. 
We restate the lemma below for the sake of presentation.
\lemMain*

\tikzstyle{decision} = [diamond, draw,
    text width=4.5em, text badly centered, node distance=3cm, inner sep=0pt]
\tikzstyle{block} = [rectangle, draw,
   text centered, rounded corners]
\tikzstyle{line} = [draw, -latex']
\tikzstyle{cloud} = [draw, ellipse,fill=red!20, node distance=3cm,
    minimum height=2em]

\begin{proof}
For notational convenience, we will replace $|S(u_i)|$ by $s_i$ and $|S(v_j)|$ by $t_j$.
We split the sum $$\sum_{i\le j}(2^{-i}u_i^T)A_s(2^{-j}v_j)$$ into several subcases depending on $i,j$ and the sizes of $S(u_i)$ and $S(v_j)$ and use the triangle inequality.  Figure \ref{fig:case_by_case} summarizes the splitting of $(i,j)$ into various terms depending on the various values of $i$, $j$, $s_i$ and $t_j$. Next, we bound each of the terms separately. By Lemma \ref{prob_lemma_2} and Corollary \ref{alternative_prob_lemma_corollary}, we know that $A_s$ satisfies the property mentioned in both of them with probability atleast $1-2e^{-3n/d^2}$. We bound the terms assuming that $A_s$ satisfies the property mentioned in Lemma \ref{prob_lemma_2} and Corollary \ref{alternative_prob_lemma_corollary}.

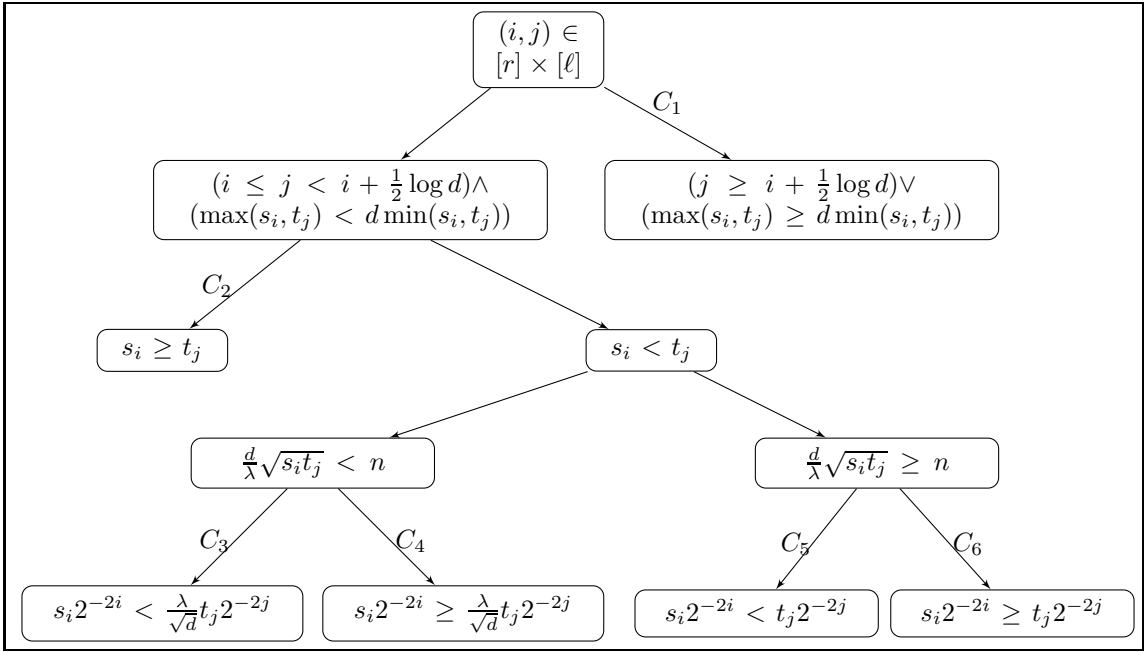
\marginbox{0pt 10pt} {\framebox{
\begin{tikzpicture}[node distance = 2cm, auto] \label{fig:case_by_case}
    \node [block, text width = 1.5cm] (init) {$(i,j)\in [r]\times [\ell]$};
    \node [block, below of = init,  xshift = -25mm, text width = 5cm]
    (non-trivial) {$(i \leq j < i + \frac{1}{2} \log d) \wedge $\\ $(\max(s_i,t_j) <
    d \min(s_i,t_j))$}; \node [block, below of = init,  xshift = +35mm, text
    width = 5cm] (trivial) {$(j \geq i + \frac{1}{2} \log d) \vee $\\
    $(\max(s_i,t_j) \geq d \min(s_i,t_j))$}; \node [block, below of =
    non-trivial , xshift = -25 mm, text width = 1.5 cm] (s_i_large) {$s_i \geq
    t_j$}; \node [block, below of = non-trivial , xshift = +4 cm, text width = 1.5 cm] (s_i_small) {$s_i < t_j$}; \node [block, below of = s_i_small, node distance = 1.5cm, xshift = -4.5 cm, text width = 3 cm] (case_1) {$\frac{d}{\lambda}\sqrt{s_i t_j} < n$}; \node [block, below of = s_i_small, node distance = 1.5cm, xshift = +30 mm, text width = 3 cm] (case_2) {$\frac{d}{\lambda}\sqrt{s_i t_j} \geq n$};
\node [block, below of = case_1,  xshift = -20 mm, text width = 3.5 cm]
(case_1_1) {$s_i 2^{-2i} < \frac{\lambda}{\sqrt{d}}t_j 2^{-2j}$}; \node [block,
below of = case_1, xshift = +20 mm, text width = 3.5 cm] (case_1_2) {$s_i
2^{-2i} \geq \frac{\lambda}{\sqrt{d}}t_j 2^{-2j}$}; \node [block, below of =
case_2, xshift = -16 mm, text width = 3 cm] (case_2_1) {$s_i 2^{-2i} < t_j 2^{-2j}$}; \node [block, below of = case_2, xshift = +18 mm, text width = 3 cm] (case_2_2) {$s_i 2^{-2i} \geq t_j 2^{-2j}$};
   \path [line] (init) -- (non-trivial) node[midway,above]{};
    \path [line] (init) -- (trivial) node[midway,above]{$C_1$};
    \path [line] (non-trivial) -- (s_i_large) node[near end,above] {$C_2$};
    \path [line] (non-trivial) -- (s_i_small) node[near end,above] {};
    \path [line] (s_i_small) -- (case_1) node[near end,above] {};
    \path [line] (s_i_small) -- (case_2) node[near end,above] {};
    \path [line] (case_1) -- (case_1_1) node[near end,above] {$C_3$};
    \path [line] (case_1) -- (case_1_2) node[near end,above] {$C_4$};
    \path [line] (case_2) -- (case_2_1) node[near end,above] {$C_5$};
    \path [line] (case_2) -- (case_2_2) node[near end,above] {$C_6$};
\end{tikzpicture}}}

\begin{claim}\label{claim:X_1_bound}
\begin{equation*}
\left|\sum_{(i,j)\in C_1} (2^{-i}u_i^T)A_s(2^{-j}v_j)\right|
\le 3\sqrt{d}\left(\sum_{i\in [r]} s_i2^{-2i} + \sum_{j\in [\ell]} t_j2^{-2j}\right).
\end{equation*}
\end{claim}
\begin{proof}
The sum is conditioned over the set of tuples $(i,j)$ in $C_1$, where
\begin{equation*}
C_1 = \left\{ (i\in [r],j\in [\ell]) \mid ( j \geq i + \frac{1}{2} \log d)\text{ or } \left(\max(s_i,t_j) \geq d \min(s_i,t_j)\right)\right\}.
\end{equation*}

By triangle inequality, 
\begin{align*}
\left|\sum\limits_{(i,j) \in C_1} 2^{-i-j} u_i^T A_s v_j \right| 
  & \leq  \left |\sum \limits_{(i,j)\in [r]\times [\ell]: j \geq i + \frac{1}{2}\log d} 2^{-i-j}u_i^TA_s v_j \right| 
+ \left|\sum_{(i,j)\in [r]\times [\ell]:i \leq j < i + \frac{1}{2}\log d,\atop \max(s_i,t_j) \geq d \min(s_i,t_j)} 2^{-i-j}u_i^TA_s v_j\right|
\end{align*}

We note that the number of edges out of any set $S$ is bounded by $d |S|$. So, $|u_i^T A_s v_j| \leq d\min(s_i,t_j)$ for any $u_i,v_j \in \{ -1,0,+1\}^n$. 
We now bound the two terms above. For the first term, 
\begin{eqnarray*}
\left |\sum \limits_{(i,j)\in [r]\times [\ell]:j \geq i + \frac{1}{2}\log d}
2^{-i-j}u_i^TA_s v_j \right|
&\leq& 
\sum_{i\in [r]} \sum_{j=i+\frac{1}{2}\log d}^{\ell} 2^{-i-j} |u_i^T A_s v_j| \\
& \leq & \sum_{i\in [r]} \sum_{j=i + \frac{1}{2}\log d}^{\ell} 2^{-i-j}d\cdot \min(s_i,t_j) \\
&\leq & \sum_{i\in [r]} \sum_{j=i + \frac{1}{2}\log d}^{\ell} 2^{-i-j}d\cdot s_i\\
&\leq& 2\sqrt{d} \sum_{i\in [r]} 2^{-2i}s_i.
\end{eqnarray*}

For the second term, 
\begin{eqnarray*}
\left|\sum_{(i,j)\in [r]\times [\ell]:i \leq j < i + \frac{1}{2}\log d,\atop \max(s_i,t_j) \geq d \min(s_i,t_j)} 2^{-i-j}u_i^TA_s v_j\right|
& \leq & \sum_{i\in [r],j\in [\ell]:i \leq j < i + \frac{1}{2}\log d,\atop \max(s_i,t_j) \geq d \min(s_i,t_j)} 2^{-i-j} |u_i^T A_s v_j|\\
& \leq & \sum_{i\in [r],j\in [\ell]:i \leq j < i + \frac{1}{2}\log d,\atop \max(s_i,t_j) \geq d \min(s_i,t_j)} 2^{-i-j} d \min(s_i,t_j) \\
& \leq & \sum_{i\in [r],j\in [\ell]:i \leq j < i + \frac{1}{2}\log d,\atop \max(s_i,t_j) \geq d \min(s_i,t_j)} 2^{-i-j} max(s_i,t_j) \\
& \leq & \sum_{i\in [r],j\in [\ell]:i \leq j < i + \frac{1}{2}\log d,\atop \max(s_i,t_j) \geq d \min(s_i,t_j)} 2^{-i-j}(s_i + t_j) \\
& = & \sum_{i\in[r]} 2^{-i} s_i \sum \limits_{j=i}^{i+\frac{1}{2} \log d} 2^{-j} + \sum_{j\in[\ell]} 2^{-j} t_j \sum\limits_{i=j-\frac{1}{2} \log d}^{i=j} 2^{-i}\\
&\leq& \frac{2}{\sqrt{d}}\sum_{i\in[r]} s_i2^{-2i} + 2\sqrt{d} \sum_{j\in [\ell]} t_j 2^{-2j}.
\end{eqnarray*}


\end{proof}

\begin{claim}\label{claim:X_2_bound}
\[
\left|\sum_{(i,j) \in C_2}(2^{-i}u_i^T)A_s(2^{-j}v_j)\right| 
\le 28 \max\left(\sqrt{d},\sqrt{\lambda\log d}\right) \sum_{i\in [r]}s_i2^{-2i}.
\]
\end{claim}
\begin{proof}
The sum is conditioned over the set of tuples $(i,j)$ in $C_2$, where

\begin{equation*}
C_2 = \{ (i,j) \in [r]\times [\ell]| (i\leq j < i+ \frac{1}{2} \log d) \text{ and }(t_j \leq s_i < d \cdot t_j) \}.
\end{equation*}
By triangle inequality the required sum is at most $\sum_{(i,j)\in C_2}2^{-i-j}|u_i^TA_sv_j|$. We note that $u_i, v_j\neq \overline{0}$ since $t_j\le s_i<dt_j$. Consider the term $|u_i^T A_s v_j|$ where $(i,j)$ is in $C_2$. We have two cases:\\

\noindent Case 1: If $ (d/\lambda) \sqrt{s_i t_j} \geq  n$, then we use Lemma \ref{prob_lemma_2} for the choice $a \leftarrow u_i, b_0 \leftarrow v_j$. This choice satisfies the conditions of Lemma \ref{prob_lemma_2}.
Hence, 
\[
|u_i^T A_s v_j|  \leq  14\sqrt{ d\cdot s_i^2  \cdot \frac{t_j}{n} \log \left( \frac{2n}{t_j} \right)} \leq  14\sqrt{d} s_i.
\]
Here, the last inequality follows by using $x\log \left(\frac{2}{x} \right) \leq 1$ for $x <1$. \\

\noindent Case 2: If $(d/\lambda) \sqrt{s_i t_j} < n$, then we use Corollary \ref{alternative_prob_lemma_corollary} for the choice $a \leftarrow v_j, b \leftarrow u_i$. This choice satisfies the conditions of Corollary \ref{alternative_prob_lemma_corollary} since $t_j\le s_i<dt_j$, condition (I) of the Lemma implies $s_i>n/d^2$, and $(d/\lambda)\sqrt{s_i t_j}<n$.
Hence, 

\begin{align*}
|u_i^T A_s v_j| & \leq  14 \sqrt{\lambda \sqrt{t_j s_i} s_i \log \left( \frac{2\cdot d \cdot t_j}{s_i} \right)}  \leq 14\sqrt{\lambda \log d} s_i.
\end{align*}
The last inequality follows since $t_j \leq s_i$. 

Thus, for $(i,j) \in C_2$, we have $|u_i^T A_s v_j| \leq 14\max(\sqrt{d},\sqrt{\lambda \log d}) s_i$. Therefore, 

\begin{align*}
\left|\sum_{(i,j) \in C_2}(2^{-i}u_i^T)A_s(2^{-j}v_j)\right| 
& \leq  \sum_{(i,j)\in C_2} 2^{-i-j} |u_i^T A_s v_j| \\
& \leq  14 \sum_{i\in[r]} \sum_{j = i}^{\infty} 2^{-i-j} \max(\sqrt{d},\sqrt{\lambda \log d}) s_i\\
& \leq  28\max{\left(\sqrt{d},\sqrt{\lambda\log d}\right)} \sum_{i\in[r]} s_i2^{-2i}.
\end{align*}
\end{proof}

\begin{claim}\label{claim:X_3_bound}
\begin{equation*}
\left|\sum_{(i,j) \in C_3}(2^{-i}u_i^T)A_s(2^{-j}v_j)\right|
=\left(\frac{\lambda}{5}+0.95 \cdot 10^{12}\sqrt{d}\right) \sum_{j\in[\ell]} t_j2^{-2j}.
\end{equation*}
\end{claim}
\begin{proof}

The sum is conditioned over the set of tuples $(i,j)$ in $C_3$, where
\begin{equation*}
C_3 = \left\{ (i,j) | (i \leq j \leq i + \frac{1}{2} \log d) \wedge (s_i \leq t_j
< d s_i) \wedge \left(\frac{d}{\lambda}\sqrt{s_it_j} < n\right) \wedge
\left(s_i2^{-2i} < \frac{\lambda}{\sqrt{d}} t_j2^{-2j}\right) \right\}.
\end{equation*}
By triangle inequality, 
\[
\left|\sum_{(i,j) \in C_3}(2^{-i}u_i^T)A_s(2^{-j}v_j)\right| 
\leq \sum \limits_{(i,j) \in C_3} 2^{-i-j} |u_i^T A_s v_j|.
\]
We note that $u_i,v_j\neq \overline{0}$ since $s_i\le t_j<ds_i$. We use Corollary \ref{alternative_prob_lemma_corollary} to bound each term $|u_i^T A_s v_j|$. We use Corollary \ref{alternative_prob_lemma_corollary} with the choice $a \leftarrow u_i$ and $b \leftarrow v_j$. This choice satisfies the conditions of Corollary \ref{alternative_prob_lemma_corollary} since $s_i\le t_j\le ds_i$, condition (I) of the Lemma implies $t_j>n/d^2$, and $(d/\lambda)\sqrt{s_it_j}<n$. 
Hence, 
\begin{align*}
\sum \limits_{(i,j) \in C_3} 2^{-i-j} |u_i^T A_s v_j|
& \leq  10\sum_{(i,j) \in C_3} 2^{-i-j}  \sqrt{\lambda \sqrt{s_i t_j} t_j \log\left(\frac{2ds_i}{t_j}\right)}\\
& < 10\sum_{(i,j) \in C_3}\frac{(\lambda)^{3/4}}{d^{1/8}} t_j 2^{-i-j} \sqrt{2^{-(j-i)}
\log\left(\frac{2\lambda\sqrt{d}}{2^{2j-2i}}\right)} \quad \quad \quad \left(\text{since } s_i 2^{-2i} <
\frac{\lambda}{\sqrt{d}} t_j 2^{-2j} \right) \\
& \leq  10\frac{(\lambda)^{3/4}}{d^{1/8}} \sum_{j\in[\ell]} t_j 2^{-2j} \sum_{ i= j-\frac{1}{2} \log d+1}^{i=j} \sqrt{2^{j-i} \log\left(\frac{2\lambda\sqrt{d}}{2^{2j-2i}}\right)} \qquad  \\
& =  90\frac{\lambda^{3/4}}{d^{1/8}} \sqrt{\sqrt{d}\log\left(
\frac{2\lambda}{\sqrt{d}}\right)} \sum_{j\in[\ell]} t_j2^{-2j} \quad \quad \quad (\text{by Lemma
\ref{agp_log_lemma} and } \lambda \geq \sqrt{d})\\
& = 90 \lambda \sqrt{\sqrt{\frac{\sqrt{d}}{\lambda}} \log\left( \frac{2\lambda}{\sqrt{d}} \right)} \sum_{j\in[\ell]} t_j 2^{-2j}.
\end{align*}

By Fact \ref{fact:one}, (we can chose an appropriate constant $c_1$) such that the above quantity is bounded by
\begin{align*}
&\left(\frac{\lambda}{5}+0.95\cdot 10^{12}\sqrt{d}\right) \sum_{j\in[\ell]} t_j2^{-2j}.
\end{align*}
\end{proof}

\begin{claim}\label{claim:X_4_bound}
\begin{equation*}
\left|\sum_{(i,j) \in C_4}(2^{-i}u_i^T)A_s(2^{-j}v_j)\right|  
=136\sqrt{d} \sum_{i \in [r]} s_i 2^{-2i}.
\end{equation*}
\end{claim}
\begin{proof}

The sum is conditioned over the set of tuples $(i,j)$ in $C_4$, where
\begin{equation*}
C_4 = \left\{(i,j) | (i \leq j < i + \frac{1}{2} \log d) \wedge (s_i \leq t_j
< d s_i) \wedge \left(\frac{d}{\lambda}\sqrt{s_it_j} < n\right) \wedge
\left(s_i2^{-2i} \geq \frac{\lambda}{\sqrt{d}} t_j2^{-2j}\right) \right\}.
\end{equation*}

By triangle inequality, 
\[
\left|\sum_{(i,j) \in C_4}(2^{-i}u_i^T)A_s(2^{-j}v_j)\right| 
\le \sum_{(i,j)\in C_4}2^{-i-j}|u_i^TA_sv_j|.
\]
We note that $u_i,v_j\neq \overline{0}$ since $s_i\le t_j<ds_i$. We use Corollary \ref{alternative_prob_lemma_corollary} to bound each term $|u_i^T A_s v_j|$. We use Corollary \ref{alternative_prob_lemma_corollary} with the choice $a \leftarrow u_i$ and $b \leftarrow v_j$. This choice satisfies the conditions of Corollary \ref{alternative_prob_lemma_corollary} since $s_i\le t_j\le ds_i$, condition (I) of the Lemma implies $t_j>n/d^2$, and $(d/\lambda)\sqrt{s_it_j}<n$. 
Hence, 
\begin{align*}
\left|\sum_{(i,j) \in C_4}(2^{-i}u_i^T)A_s(2^{-j}v_j)\right| 
&\le \sum_{(i,j)\in C_4}2^{-i-j}|u_i^TA_sv_j|\\
& \leq  10 \sum_{(i,j) \in C_4} 2^{-i-j}  \sqrt{\lambda \sqrt{s_i t_j} t_j \log\left(\frac{2ds_i}{t_j}\right)}\\
& = 10\sum_{(i,j) \in C_4} 2^{-i-j} \sqrt{\lambda} s_i \sqrt{\left(\frac{t_j}{s_i}\right)^{\frac{3}{2}} \log\left( \frac{2 d s_i}{t_j} \right)}\\
& \le  10\sum_{(i,j) \in C_4} 2^{-i-j} \frac{d^{3/8}}{\lambda^{1/4}}s_i
\sqrt{2^{3j-3i} \log\left(\frac{2\lambda \sqrt{d}}{2^{2j-2i}}\right)} 
\end{align*}
Above we use the fact that $x^{\frac{3}{2}} \log\left(\frac{c}{x}\right)$ is an increasing function if $x \leq \frac{c}{2}$ and $s_i2^{-2j} \geq \frac{\lambda}{\sqrt{d}} t_j 2^{-2j}$. Therefore,
\begin{align*}
\left|\sum_{(i,j) \in C_4}(2^{-i}u_i^T)A_s(2^{-j}v_j)\right| 
& \leq  10\sum_{i \in [r]} \frac{d^{3/8}}{\lambda^{1/4}}s_i 2^{-2i} \sum_{j = i}^{j= i+ \frac{1}{2}\log d-1} \sqrt{2^{j-i} 2 \log\left(\frac{2\lambda \sqrt{d}}{2^{2j-2i}}\right)} \qquad \\
& =  90 \sum_{i \in [r]} \frac{d^{3/8}}{\lambda^{1/4}}s_i 2^{-2i}  \sqrt{\sqrt{d}
\log\left(\frac{2 \lambda}{\sqrt{d}}\right)} && (\text{by Lemma
\ref{agp_log_lemma}})\\
& =  90\sum_{i \in [r]} d^{\frac{1}{2}} s_i 2^{-2i}  \sqrt{\sqrt{\frac{\sqrt{d}}{\lambda}} \log\left(\frac{2\lambda}{\sqrt{d}}\right)}\\
&= 136\sqrt{d}\sum_{i \in [r]} s_i 2^{-2i}.
\end{align*}
That last equality is because, $\lambda\ge \sqrt{d}$ for every $d$-regular graph and hence $\sqrt{\frac{\sqrt{d}}{\lambda}} \log\left( \frac{2 \lambda}{\sqrt{d}}\right) \leq 1.502$.
\end{proof}

\begin{claim}\label{claim:X_5_bound}
\begin{equation*}
\left|\sum_{(i,j) \in C_5}(2^{-i}u_i^T)A_s(2^{-j}v_j)\right|  
\le 56\sqrt{d}\left(\sum_{j \in [l]} t_j2^{-2j} + \sum_{i \in [r]} s_i2^{-2i} \right).
\end{equation*}
\end{claim}
\begin{proof}
The sum is conditioned over the set of tuples $(i,j)$ in $C_5$, where
\begin{equation*}
C_5 = \left\{ (i,j) | (i \leq j < i + \frac{1}{2} \log d) \wedge (s_i \leq t_j
< d s_i) \wedge \left(\frac{d}{\lambda}\sqrt{s_it_j} \geq n\right) \wedge
\left(s_i2^{-2i} <  t_j2^{-2j}\right) \right\}.
\end{equation*}

By triangle inequality,
\[
\left|\sum_{(i,j) \in C_5}(2^{-i}u_i^T)A_s(2^{-j}v_j)\right|  
 \leq \sum \limits_{j\in[\ell]:\exists i\in [r]\text{ with }(i,j)\in C_5}2^{-2j} \left|\sum \limits_{i:(i,j) \in C_5} 2^{-i+j} u_i^T A_s v_j\right|.
\]
We note that $u_i,v_j\neq \overline{0}$ since $s_i\le t_j<ds_i$ for every $(i,j)\in C_5$. Let us fix $j$ such that there exists $(i,j)\in C_5$. We bound 
\[
\left|\sum \limits_{i\in \{j-(1/2)\log{d},\ldots, j\}:\atop (i,j) \in C_5} 2^{-i+j} u_i^T A_s v_j\right|
\]
using Lemma \ref{prob_lemma_2}. 
We will use Lemma \ref{prob_lemma_2} for the choice $a \leftarrow v_j$ and for every $k=0,1,\ldots, (1/2)\log{d}$, we take $b_k \leftarrow u_{j-k}$ if $(j-k,j)\in C_5$ and $b_k\leftarrow \overline{0}$ if $(j-k,j)\not\in C_5$.  This choice satisfies the conditions of Lemma \ref{prob_lemma_2} since (i) condition (I) of the Lemma implies $S(b_k)$ are mutually non-intersecting, (ii) $|S(v_j)|=t_j\ge 2^{2j-2i} s_i=2^{2j-2i}|S(u_{i})|$ for every $(i,j)\in C_5$ implies $|S(a)|\ge 2^{2k}|S(b_k)|$ for every $k=0,1,\ldots, (1/2)\log{d}$, and (iii) $b_k$ is non-zero if and only if $(j-k,j)\in C_5$ implies $(d/\lambda)\sqrt{|S(b_k)||S(a)|}\ge n$ for every non-zero $b_k$. 
Hence, by Lemma \ref{prob_lemma_2}, we have 
\begin{align*}
\sum \limits_{j\in[\ell]}2^{-2j} \left|\sum \limits_{i:(i,j) \in C_5} 2^{-i+j} u_i^T A_s v_j\right|
& \leq 14 \sum_{j\in[\ell]} 2^{-2j} \sqrt{\frac{d t_j^2}{n} \sum_{i=j-\frac{1}{2}\log d}^{i=j} s_i2^{-2i+2j} \log\left(\frac{2n}{t_j}\right)} \\
& =   14\sqrt{d} \sum_{j\in[\ell]} \sqrt{2^{-2j} \frac{t_j^2}{n} \log\left(\frac{2n}{t_j}\right)\sum_{i=j-\frac{1}{2}\log d}^{i=j} s_i2^{-2i} }. \\
\end{align*}

Next, we group $v_j$ according to their support sizes and then sum them together. For $c = 0,1,2,\dots,\log(n)$, let $J_c$ be the set of indices $j\in[\ell]$ s.t. $n/2^c \leq t_j < 2n/2^c$ and for non-empty sets $J_c$, define $j_c := \min(j\in J_c)$. With this notation, the above sum is
\begin{align*}
&\le 14\sqrt{d}  \sum_{c=0}^{\log{n}} \sum_{j \in J_c}\sqrt{4 n2^{-2j-2c}\log(2\cdot2^c) \sum_{i=j-1/2\log d+1}^{i=j} s_i2^{-2i} }
&& \left(\frac{n}{2^c} \leq t_j < \frac{2n}{2^c}\right) \\
& \leq  14\sqrt{d} \sum_{c=0}^{\log{n}} \sum_{j \in J_c} \frac{1}{2}\left(4
n2^{-j-j_c-c} + 2^{-j+j_c-c}\log(2\cdot2^c)\sum_{i=j-1/2\log d+1}^{i=j}
s_i2^{-2i}\right) && (G.M.\le A.M.)\\
& =  28\sqrt{d} \sum_{c=0}^{\log{n}} \sum_{j \in J_c}  n2^{-j-j_c-c} 
+ 7\sqrt{d} \sum_{c=0}^{\log{n}} \sum_{j \in J_c} \sum_{i=j-1/2\log d+1}^{i=j} 2^{-j+j_c-c}\log(2\cdot2^c) s_i2^{-2i} \\
& \leq  28\sqrt{d} \sum_{c=0}^{\log{n}} \sum_{j \in J_c} \frac{n}{2^{c}} 2^{-j-j_c} +  7\sqrt{d}\sum_{i\in[r]} s_i2^{-2i} \sum_{c=0}^{\log{n}} \frac{\log(2\cdot2^c)}{2^c} \sum_{j \in J_c} 2^{-j+j_c}.\\
\end{align*}

We observe that
\[
\sum_{c=0}^{\log{n}} \sum_{j\in J_c}\frac{n}{2^c}2^{-j-j_c}\le \sum_{c=0}^{\log{n}}\sum_{j\in J_c}t_j2^{-j-j_c}\le \sum_{c=0}^{\log{n}}\sum_{j\in J_c}t_j2^{-2j}=\sum_{j\in[\ell]}t_j2^{-2j}.
\]

Moreover, $\sum \limits_{j\in J_c} 2^{-j+j_c} \leq 2$ and $\sum_{c=0}^{\log{n}} \frac{\log(2.2^c)}{2^c} \leq 4$. Substituting these we have the claim.
\end{proof}

\begin{claim}\label{claim:X_6_bound}
\begin{equation*}
\left|\sum_{(i,j) \in C_6}(2^{-i}u_i^T)A_s(2^{-j}v_j)\right| 
=154\sqrt{d} \sum_{i\in[r]} s_i 2^{-2i}.
\end{equation*}
\end{claim}
\begin{proof}
The sum is conditioned over the set of tuples $(i,j)$ in $C_6$, where
\begin{equation*}
C_{6} =\left\{ (i,j)| (i \leq j \leq i + \frac{1}{2} \log d) \wedge (s_i \leq t_j
< d s_i) \wedge (\frac{d}{\lambda}\sqrt{s_it_j} \geq n) \wedge
\left(s_i2^{-2i} \geq  t_j2^{-2j}\right) \right\}.
\end{equation*}
By triangle inequality, 
\[\left|\sum_{(i,j) \in C_6}(2^{-i}u_i^T)A_s(2^{-j}v_j)\right|
\leq  \sum_{(i,j)\in C_6} 2^{-i-j}| u_i^T A_s v_j|. \]
We will use Lemma \ref{prob_lemma_2} to bound each term $|u_i^TA_s v_j|$. We use Lemma \ref{prob_lemma_2} with the choice $a\leftarrow v_j, b_0 \leftarrow u_i$. This choice satisfies the conditions of Lemma \ref{prob_lemma_2} since $s_i\le t_j<ds_i$ and $(d/\lambda)\sqrt{s_it_j}\ge n$. Hence,
\begin{align*}
\sum_{(i,j)\in C_6} 2^{-i-j}| u_i^T A_s v_j| 
 \leq  14 \sum_{(i,j) \in C_6} 2^{-i-j} \sqrt{\frac{d s_i t_j^2}{n} \log\left(\frac{2n}{t_j}\right)}. 
\end{align*}
Next, we divide the tuples in $C_{6}$ into two parts depending on the value of $i$ and $j$:
\begin{align*}
C_6' & :=  \{ (i,j) | (i,j) \in C_6, (i \leq j < i+ \frac{1}{2}\log(n/s_i)) \}\text{ and }  \\
C_6'' & :=  \{ (i,j) | (i,j) \in C_6, (j \geq i+ \frac{1}{2}\log(n/s_i)) \}.
\end{align*}

Let us consider the above RHS sum over tuples $(i,j)$ in $C_6'$. 
\begin{align*}
14 \sum_{(i,j) \in C_6'} 2^{-i-j} \sqrt{\frac{d s_i t_j^2}{n} \log\left(\frac{2n}{t_j}\right)}
& = 14\sqrt{d}\sum_{(i,j) \in C_6'} 2^{-2i} s_i \sqrt{2^{-2j+2i}\frac{ 1}{n s_i } t_j^2 \log\left(\frac{2n}{t_j}\right)} \\
&\leq 14\sqrt{d}\sum_{(i,j) \in C_6'}s_i 2^{-2i}  \sqrt{\frac{s_i  2^{2j-2i}}{n} \log\left(\frac{2n}{s_i2^{2j-2i}}\right)} \quad \quad \quad (t_j 2^{-2j} \leq s_i 2^{-2i})\\
&\leq 14\sqrt{d}\sum_{i\in[r]}s_i 2^{-2i} \sum_{j=i}^{j=i+ \frac{1}{2}\log(\frac{n}{s_i})} \sqrt{\frac{s_i  2^{2j-2i}}{n} \log\left(\frac{2n}{s_i2^{2j-2i}}\right)}\\
&\le 126\sqrt{d} \sum_{i\in[r]} s_i2^{-2i}.
\end{align*}
In the above, the last inequality is by using Lemma \ref{agp_log_lemma} for $\sum_{j=i}^{j=i+ \frac{1}{2}\log(\frac{n}{s_i})} \sqrt{\frac{s_i  2^{2j-2i}}{n} \log\left(\frac{2n}{s_i2^{2j-2i}}\right)}$. 
Next, let us consider the RHS sum over tuples $(i,j)$ in $C_6''$.
\begin{align*}
14 \sum_{(i,j) \in C_6''} 2^{-i-j} \sqrt{\frac{d s_i t_j^2}{n} \log\left(\frac{2n}{t_j}\right)} 
& = 14\sqrt{d} \sum_{(i,j) \in C_6''} s_i 2^{-i-j} \sqrt{\frac{t_j}{s_i}} \sqrt{\frac{ t_j}{n} \log\left(\frac{2n}{t_j}\right)}\\
& \leq 14\sqrt{d}) \sum_{(i,j) \in C_6''} s_i 2^{-i-j} \sqrt{\frac{n}{s_i}} && \left(t_j \leq n, x \log \frac{2}{x} \leq 1 \right)\\
& \leq  14\sqrt{d}\sum_{i\in[r]} s_i 2^{-2i} \sum_{j=i+ \frac{1}{2}\log(n/s_i)}^{\infty} 2^{-j+i} \sqrt{\frac{n}{s_i} }\\
&\leq   28\sqrt{d}\sum_{i\in[r]} s_i 2^{-2i}.
\end{align*}
The claim follows from the above two bounds. 
\end{proof}

We now obtain the required bound for conclusion 1 of the Lemma from Claims \ref{claim:X_1_bound}, \ref{claim:X_2_bound}, \ref{claim:X_3_bound}, \ref{claim:X_4_bound}, \ref{claim:X_5_bound}, and \ref{claim:X_6_bound}:
\begin{eqnarray*}
\left|\sum_{(i,j) \in [r]\times[\ell]}(2^{-i}u_i^T)A_s(2^{-j}v_j)\right| 
\le 377\max\left(\sqrt{\lambda \log d},\sqrt{d}\right) \sum_{i\in[r]} s_i2^{-2i} + \left(\frac{\lambda}{5} + 10^{12}\sqrt{d}\right)\sum_{j\in[\ell]} t_j 2^{-2j}.
\end{eqnarray*}

For conclusion 2 of the Lemma, we observe that if $s_i\ge t_j$ for all $i\in[r], j\in [\ell]$, then $C_3, C_4, C_5, C_6$ are empty. Thus the bound follows from Claims  \ref{claim:X_1_bound} and \ref{claim:X_2_bound}:
\begin{eqnarray*}
\left|\sum_{(i,j) \in [r]\times[\ell]}(2^{-i}u_i^T)A_s(2^{-j}v_j)\right|
&\leq& 31\max\left(\sqrt{\lambda \log d},\sqrt{d}\right) \left(\sum_{i \in [r]} s_i2^{-2i} + \sum_{j \in [l]} t_j 2^{-2j}\right).
\end{eqnarray*}
\end{proof}

\section{Proof of Theorem \ref{thm:main2}}\label{proof-of-random-shift-lifts}
To prove Theorem \ref{thm:main2}, we need the following modified version of Lemma \ref{main_lemma}.

\begin{lemma}\label{modified_main_lemma}
Let $G$ be a $d$ -regular graph with non-trivial eigenvalues at most $\lambda$ in absolute value $\lambda\ge\sqrt{d}$, $2 \le d \leq \sqrt{n/3 \ln n}$ and let $A$ be the adjacency matrix of $G$. Let $A'$ be a random $n \times n$ real matrix whose entries $A'(i,j)$ are random variables with mean $0$, $|A'(i,j)| \leq A(i,j)$ for all $i,j$, and the entries $A'(i,j)$ are independent from all other entries except $A'(j,i)$. There exist constants $c_1, c_2\ge 1000, c_3, c_4$ such that the following statements hold with probability at least $1-e^{-(n/d^2)}$ (over the random choice of $A'$). 
\begin{enumerate}
\item For all $u_1,u_2, \ldots u_r \in \{0, \pm 1,\pm \frac{1}{2}\}^n$, $v_1,v_2 \ldots,v_{\ell} \in \{0,\pm 1,\pm \frac{1}{2}\}^n$ satisfying

\begin{itemize}
\item[(I)] $S(u_i) \cap S(u_j) = \phi$ for every $i,j \in [r]$ and $S(v_i) \cap S(v_j) = \phi$ for every $i,j \in [\ell]$, and
\item[(II)] Either $|S(u_i)| > n/d^2$ for every $i \in [r]$ with non-zero $u_i$, or $|S(v_i)| > n/d^2$ for every $i \in [\ell]$ with non-zero $v_i$,
\end{itemize}
we have 
\begin{equation*}
\left |\sum \limits_{i \leq j} (2^{-i}u_i^T) A' (2^{-j}v_j)\right| 
\le c_1\max\left(\sqrt{\lambda \log d},\sqrt{d}\right)
\sum_{i=1}^r |S(u_i)| 2^{-2i} +
\left(\frac{\lambda}{c_2} + c_3\sqrt{d}\right)\sum_{j=1}^{\ell}|S(v_j)| 2^{-2j}
\end{equation*}
\item For all $u_1,u_2, \ldots u_r \in \{0, \pm 1,\pm \frac{1}{2}\}^n$, $v_1,v_2 \ldots,v_{\ell} \in \{0,\pm 1,\pm \frac{1}{2}\}^n$ satisfying $(I),(II)$ and 
\begin{itemize}
\item[(III)] $|S(u_i)| > |S(v_j)|$ for every $i \in [r],j\in [\ell]$ with non-zero $u_i$,
\end{itemize}
we have 
\begin{equation*}
\left|\sum\limits_{i \leq j} (2^{-i}u_i^T) A' (2^{-j}v_j)\right| \leq c_4\max\left(\sqrt{\lambda \log d},\sqrt{d}\right)
\left(\sum_{i=1}^r |S(u_i)| 2^{-2i} + \sum_{j=1}^{\ell} |S(v_j)| 2^{-2j}\right).
\end{equation*}
\end{enumerate}
\end{lemma}

The proof of Lemma \ref{modified_main_lemma} is identical to that of Lemma \ref{main_lemma}. In the proof of Lemma \ref{main_lemma}, we used the concentration inequalities from Lemma \ref{prob_lemma_2} and Corollary \ref{alternative_prob_lemma_corollary}. We note that these concentration inequalities were obtained using Hoeffding's inequality. Since Hoeffding's inequality is applicable when the random variables are bounded, we have the version of Lemma \ref{prob_lemma_2} and Corollary \ref{alternative_prob_lemma_corollary} applicable to the random matrix $A'$. As a consequence, we obtain Lemma \ref{modified_main_lemma} by following the same proof as that of Lemma \ref{main_lemma}. We avoid repeating the proof for brevity. 
 

\begin{proof}[Proof of Theorem \ref{thm:main2}]
The proof is very similar to the proof of Theorem \ref{main_theorem_easy}. However, in order to avoid a loss of factor $4$, we avoid discretizing in the first step, but discretize only for certain cases. Using Lemma \ref{lem:shiftspectrum}, we know that for a shift $k$-lift, $\lambda_{new}$ is the maximum absolute value in the set  
\[
\bigcup_{\omega:\ \omega \text{ is a $k$-th primitive root of unity},\ \omega\neq 1} \text{eigenvalues}\left(A_s(\omega)\right).
\] 
We will bound the probability that the maximum eigenvalue of $A_s(\omega)$ is large for $\omega$ being a fixed primitive $k$-th root of unity. A union bound over the $k-1$ primitive $k$-th roots of unity bounds the maximum eigenvalues of all $k-1$ matrices simultaneously. 


Let us fix $\omega$ to be a primitive $k$-th root of unity and bound the eigenvalues of $A_s(\omega)$. We need to bound $\max_{x \in \C^n} |x^* A_s(\omega) x/x^* x|$ where $x^*$ denotes the complex conjugate of vector $x$. Let $x = q + i w \in \C^n$ where $q, w \in \R^n$.  We consider a decomposition of $q,w$ (similar to but not the same as the diadic decomposition) into a sequence of vectors $y_i$'s and $z_i$'s for $i=0,1,\ldots$ respectively as follows:
\[
[y_i]_j := \begin{cases}
	q_j & \text{if  } 2^{-i-1} < |q_j| \leq 2^{-i},\\
    	0 & \text{otherwise},
\end{cases}
\]
\[
[z_i]_j := \begin{cases}
	w_j & \text{if  } 2^{-i-1} < |w_j| \leq 2^{-i},\\
    	0 & \text{otherwise}.
\end{cases}
\]
Let us partition the set of indices $\{0,1,\ldots\}$ into two sets $M_r:=\{i:|S(y_i)|<n/d^2\}$ and $L_r:=\{i:|S(y_i)|\ge n/d^2\}$ and define $y_{M_r} := \sum_{i \in M_r} y_i$ and $y_{L_r} := \sum_{i \in L_r} y_i$. Similarly, define $M_c$ and $L_c$ based on the support of $z_i$'s and define $z_{M_c}$ and $z_{L_c}$. We will refer to vectors $y_{M_r}, z_{M_c}$ as ``type M'' vectors, and $y_{L_r}$ and $z_{L_c}$ as ``type L'' vectors. We note that 
\[
x^*x=\|y_{M_r}\|^2 + \|y_{L_r}\|^2 + \|z_{M_c}\|^2 + \|z_{L_c}\|^2. 
\]
By splitting the terms in $|x^* A_s(\omega) x|$, we get
\begin{align}\label{eq:splitting}
|x^* A_s(\omega) x| 
\leq |(y_{M_r}+ i z_{M_c})^* A_s(\omega) (y_{M_r} + i z_{M_c})| + |z_{L_c}^T A_s(\omega) y_{L_r}| + |y_{L_r}^TA_s(\omega) z_{L_c}| \notag \\
 + |y_{L_r}^T A_s(\omega) y_{L_r}| + |y_{L_r}^T A_s(\omega)y_{M_r}| +  |y_{M_r}^T A_s(\omega)y_{L_r}| \notag \\
+ |z_{L_c}^T A_s(\omega) z_{L_c}| + |z_{L_c}^T A_s(\omega)z_{M_c}| +  |z_{M_c}^T A_s(\omega)z_{L_c}| \notag \\
 +  |y_{L_r}^T A_s(\omega) z_{M_c}| + |z_{M_c}^T A_s(\omega) y_{L_r}| + |z_{L_c}^T A_s(\omega) y_{M_r}| + |y_{M_r}^T A_s(\omega) z_{L_c}|.
\end{align}

To derive an upper bound on $|x^* A_s(\omega) x |$, we will show upper bounds for each of the terms in the RHS using Lemma \ref{modified_main_lemma}. We note that the concentration inequalities given in parts 1 and 2 of Lemma \ref{modified_main_lemma} hold with probability at least $1-e^{-n/d^2}$ for some constants $c_1, c_2\ge 1000, c_3, c_4$. Assuming parts 1 and 2 of Lemma \ref{modified_main_lemma}, we have the following claims:

\begin{claim}\label{claim:generalized_signing_bound_1}
\begin{equation*}
|(y_{M_r} + i z_{M_c})^* A_s(\omega) (y_{M_r} + i z_{M_c})|\leq \left( \lambda + \frac{128}{d}\right)||y_{M_r} + i z_{M_c}||^2.
\end{equation*}
\end{claim}

\begin{claim}\label{claim:generalized_signing_bound_2}
For any type L vectors $a$ and $b$,
\begin{equation*} 
|a^T A_s(\omega) b| \leq \left(\frac{32\lambda}{c_2} +32(c_1+c_3)\left(\max\left(\sqrt{\lambda \log(d)}, \sqrt{d}\right)\right)\right)\left(||a||^2 + ||b||^2\right).
\end{equation*}
\end{claim}

\begin{claim}\label{claim:generalized_signing_bound_3}
For any vector $a$ of type $M$ and vector $b$ of type $L$,
\begin{equation*} 
|a^T A_s(\omega) b| \leq \frac{32\lambda}{c_2}||b||^2 + 32(c_1+c_3+c_4)\left( \max\left(\sqrt{\lambda \log(d)}, \sqrt{d}\right)\right)\left(||b||^2 + ||a||^2\right).
\end{equation*}
\end{claim}

We note that all terms in the RHS of inequality (\ref{eq:splitting}) fall into one of the three categories given in Claims \ref{claim:generalized_signing_bound_1}, \ref{claim:generalized_signing_bound_2} and \ref{claim:generalized_signing_bound_3} above. Using these bounds, the following holds with probability at least $1-e^{-(n/d^2)}$:
\begin{align*}
\left|x^* A_s(\omega) x\right| 
& \leq  \left(\lambda + \frac{128}{d}\right)\|y_{M_r} + i z_{M_c}\|^2 + \frac{256\lambda}{c_2}\left(\|y_{L_r}\|^2 + \|z_{L_c}\|^2\right)  \\
& \quad \quad + 256(c_1+c_3+c_4)\left(\max\left(\sqrt{\lambda \log(d)}, \sqrt{d}\right)\right)\left(\|y_{M_r}\|^2 + \|z_{M_c}\|^2 + \|y_{L_r}\|^2 + \|z_{L_c}\|^2 \right) \\
& \leq  \left(\lambda + 288(c_1+c_3+c_4)\left( \max\left(\sqrt{\lambda \log(d)} , \sqrt{d}\right) \right)\right) x^* x. 
\end{align*}
The last inequality is because $c_2\ge 1000$ and $d\ge 2$. Taking a union bound over the $k-1$ primitive roots of unity shows that there exists a constant $c$ such that with probability at least $1-ke^{-(n/d^2)}$, all new eigenvalues of a random shift $k$-lift have absolute value at most 
\[
\lambda + c \max\left(\sqrt{\lambda \log(d)} , \sqrt{d}\right) .
\]
\end{proof}

\begin{proof}[Proof of Claim \ref{claim:generalized_signing_bound_1}] We observe that $|(y_{M_r} + i z_{M_c})^* A_s(\omega) (y_{M_r} + i z_{M_c})|  \leq y'^T A y'$ where $y'$ is a real vector whose $j$-th coordinate is equal to the absolute value of the $j$-th element in $y_{M_r} + i z_{M_c}$ and $A$ is the adjacency matrix of the base graph.
Let $J=vv^T$ and $J'=v'v'^T$ where $v$ is the all ones vector and $v'$ is defined as $v_i'=1$ for $i\in\{1,\ldots, n/2\}$ and $v_i'=-1$ for $i\in \{n/2+1,\ldots, n\}$. For non-bipartite graph $G$, we have 
\begin{align*}
y'^T A y' &= y'^T (A - \frac{d}{n}J) y' + y'^T \left(\frac{d}{n}J\right) y'
\le \lambda \|y'\|^2 + y'^T \left(\frac{d}{n}J\right) y' \\
&= \lambda \|(y_{M_r} + i z_{M_c})\|^2+ y'^T \left(\frac{d}{n}J\right) y'. 
\end{align*}
Above, we have used the fact that the maximum eigenvalue of $A- (\frac{d}{n} J)$ is $\lambda$.
Similarly, for bipartite graphs, we have
\begin{align*}
y'^T A y' &= y'^T (A - \frac{d}{n}J+\frac{d}{n}J') y' + y'^T \left(\frac{d}{n}J\right) y'-y'^T\left(\frac{d}{n}J'\right)y'\\
&\le \lambda \|y'\|^2 + y'^T \left(\frac{d}{n}J\right) y' - y'^T \left(\frac{d}{n}J'\right) y'\\
&\le \lambda \|(y_{M_r} + i z_{M_c})\|^2+ y'^T \left(\frac{2d}{n}J\right) y'. 
\end{align*}

It remains to bound $|y'^T \frac{d}{n} J y'|$. 
Let $y_{M_r}'$ and $z_{M_c}'$ be vectors obtained by taking the absolute values of the coordinates of $y_{M_r}$ and $z_{M_c}$ respectively. We have 
\begin{align*}
y'^T \left(\frac{d}{n} J\right) y'&\leq (y_{M_r}' + z_{M_c}')^T \left(\frac{d}{n} J\right) (y_{M_r}' + z_{M_c}').
\end{align*} 
We recall that the number of entries between $2^{-i-1}$ and $2^{-i}$ in $y_{M_r}'$ and $z_{M_c}'$ are less than $\frac{n}{d^2}$. We will show that $|u^T (\frac{d}{n}) J v| \leq \frac{4}{d}(\|u\|^2 + \|v\|^2)$ where $u,v \in \{y_{M_r}', z_{M_c}' \}$. 

Let $u,v\in \{y_{M_r}', z_{M_c}' \}$. By Lemma \ref{DiscretizationLemma},  there exist $u'$, $v'$
s.t. $|u^T \frac{d}{n}J v| \leq |u'^T \frac{d}{n} J v'|$ where $u',v' \in
\{0,\pm \frac{1}{2}, \pm \frac{1}{4}, \ldots \}^n$, $\|u'\|^2 \leq 4\|u\|^2$, and
$\|v'\|^2 \leq 4 \|v\|^2$. Consider the diadic decomposition of $u' = \sum_{i=0}^{\infty}2^{-i} u_i$ obtained as follows: a coordinate of $u_i$ is $1$ if the corresponding coordinate of $u'$ is $2^{-i}$, it is $-1$ if the corresponding coordinate of $u'$ is $-2^{-i}$ and is $0$ otherwise. Similarly, define the diadic decomposition of $v'=\sum_{j=0}^{\infty} 2^{-j} v_j$.  We note that all entries between $2^{-i-1}$ and $2^{-i}$ in $u$ and $v$ are rounded to either $2^{-i-1}$ or $2^{-i}$ in $u'$ and $v'$ and all entries between $-2^{-i-1}$ and $-2^{i}$ are rounded to either $-2^{-i-1}$ or $-2^{-i}$. Since the number of entries in $u,v$ with absolute value between $2^{-i-1}$ and $2^{-i}$ is at most $n/d^2$, we get $|S(u_i)|,|S(v_j)| < \frac{2n}{d^2}$ for all $i,j$. Thus, 
\begin{eqnarray*}
\left|u' \left(\frac{d}{n} J \right) v'\right| & = &  \left|\sum_{i,j=0}^{\infty} 2^{-i-j} u_i^T \left(\frac{d}{n} J\right) v_j\right| \\
& \leq & \sum_{i = 0}^{\infty} \sum_{j=i}^{\infty} 2^{-i-j} \frac{d}{n} |u_i^T J v_j| + \sum_{j = 0}^{\infty} \sum_{i = j+1}^{\infty} 2^{-i-j} \frac{d}{n} |u_i^T J v_j| \\
& \leq & \sum_{i = 0}^{\infty} \sum_{j=i}^{\infty} 2^{-i-j} \frac{d |S(u_i)| |S(v_j)|}{n}+\sum_{j= 0}^{\infty} \sum_{i=j+1}^{\infty} 2^{-i-j} \frac{d |S(v_j)| |S(u_i)|}{n} \\
& \leq & 2\sum_{i=0}^{\infty} 2^{-2i} \frac{|S(u_i)|}{d} \sum_{j = i}^{\infty} 2^{-j+i} + 2\sum_{j=0}^{\infty} 2^{-2j} \frac{|S(v_j)|}{d} \sum_{i = j+1}^{\infty} 2^{-i+j} \\
& \leq & \frac{4 }{d} \left(\sum_{i=0}^{\infty} |S(u_i)| 2^{-2i} + \sum_{j=0}^{\infty} |S(v_j)|2^{-2j}\right)\\
& \leq & \frac{4}{d} \left(\|u'\|^2 + \|v'\|^2\right).
\end{eqnarray*}

For $u,v \in \{y'_{M_r}, z'_{M_c} \}$, $|u^T (\frac{d}{n}J) v| \leq |u'^T (\frac{d}{n}J) v'| \leq \frac{4}{d} (\|u'\|^2 + \|v'\|^2) \leq \frac{16}{d}(\|u\|^2 + \|v\|^2)$. Therefore, 
\begin{eqnarray*}
y'^T \left(\frac{d}{n} J\right) y' &\leq & (y_{M_r}' + z_{M_c}')^T \left(\frac{d}{n} J\right) (y_{M_r}' + z_{M_c}') \\
& \leq & y_{M_r}'^T  \left(\frac{d}{n} J\right) y_{M_r}' + y_{M_r}'^T  \left(\frac{d}{n} J\right) z_{M_c}' + z_{M_c}'^T  \left(\frac{d}{n} J\right) y_{M_r}' + z_{M_c}'^T  \left(\frac{d}{n} J\right) z_{M_c}'\\
& \le& \left(\frac{16}{d}\right)\left(\|y_{M_r}'\|^2 + \|y_{M_r}'\|^2  + \|y_{M_r}'\|^2 + \|z_{M_c}'\|^2 + \|z_{M_c}'\| + \|y_{M_r}'\|^2 + \|z_{M_c}'\|+\|z_{M_c}'\|\right)\\
& =& \left(\frac{64}{d}\right)\left(\|y_{M_r}'\|^2 + \|z_{M_c}'\|^2\right) 
=  \left(\frac{64}{d}\right) \left( \|y_{M_r}\|^2 + \|z_{M_c}\|^2\right) 
= \left(\frac{64}{d}\right)\|y_{M_r} + i z_{M_c}\|^2
\end{eqnarray*}

Thus, we have $|(y_{M_r} + i z_{M_c})^* A_s(\omega) (y_{M_r} + i z_{M_c})| \leq y' A y' \leq (\lambda + (128/d)) \|(y_{M_r} + i z_{M_c})\|^2$.

\end{proof}

In order to show Claims \ref{claim:generalized_signing_bound_2} and \ref{claim:generalized_signing_bound_3}, we divide the matrix into its real and imaginary part: $A_s(\omega) = A_s^1(\omega) + i A_s^2(\omega)$ where $A_s^1(\omega)$ and $A_s^2(\omega)$ are real matrices. For any two vectors $a,b\in \R^n$,
\begin{equation*}
|a^T A_s(\omega) b| \leq |a^T A_s^1(\omega) b| + |a^T A_s^2(\omega) b|.
\end{equation*}

We will bound $|a^T A_s'(\omega) b|$ where $A_s'(\omega) \in \{A_s^1(\omega),A_s^2(\omega)\}$ for $a,b$ as in Claims \ref{claim:generalized_signing_bound_2} and \ref{claim:generalized_signing_bound_3}. We start by discretizing $a$ and $b$.  By Lemma \ref{DiscretizationLemma},
there exist $a',b'$ such that $|a^T A_s'(\omega)b| \leq |a'^T A_s'(\omega) b'|$ where
$a',b' \in \{0,\pm\frac{1}{2}, \pm\frac{1}{4} \dots \}^n$ and $\|a'\|^2 \leq 4\|a\|^2$ and $\|b'\|^2 \leq 4\|b\|^2$. Moreover, every entry of $a$ and $b$ between 
$2^{-i-1}$ and $2^{-i}$ is rounded to either $2^{-i-1}$ or $2^{-i}$ in $a'$ and $b'$ respectively (similarly, every entry between $-2^{-i-1}$ and $-2^{-i}$ is rounded to either $-2^{-i-1}$ or $-2^{-i}$). 
Consider the following vectors $\{u_i\}_{i\in \{0,1,\ldots\}}$, $\{v_i\}_{i\in \{0,1,\ldots\}}$ obtained from $a'$, $a$ and $b$, $b'$ respectively:
\[
[u_i]_j := \begin{cases}
	2^ia'_j, & \text{if  } 2^{-i-1} \leq |a_j|  < 2^{-i}\\
    	0, & \text{otherwise}
\end{cases}
\]
\[
[v_i]_j := \begin{cases}
	2^ib'_j, & \text{if  } 2^{-i-1} \leq |b_j| < 2^{-i}\\
    	0, & \text{otherwise}
\end{cases}
\]
We observe that  $u_i, v_i \in \{0,\pm \frac{1}{2}, \pm 1 \}^n$, $|a'^T A_s'(\omega) b'| = |\sum_{i,j=0}^{\infty}2^{-i-j} u_i^T A_s'(\omega) v_j|$,  $\|a'\|^2 = \sum_i 2^{-2i} \|u_i\|^2 \geq \frac{1}{4} \sum_i 2^{-2i} |S(u_i)|$ and
\begin{eqnarray*}
\left|\sum_{i,j=0}^{\infty} 2^{-i-j} u_i^T A_s'(\omega) v_j\right| & \leq & \left|\sum_{i\leq j} 2^{-i-j} u_i^T A_s'(\omega) v_j\right| + \left|\sum_{i<j} 2^{-i-j} v_i^T A_s'(\omega) u_j\right|.
\end{eqnarray*}

\begin{proof}[Proof of Claim \ref{claim:generalized_signing_bound_2}]
Since $a$ and $b$ are type $L$ vectors, we have $|S(u_i)|,|S(v_j)| \geq \frac{n}{d^2}$ for all non-zero $u_i,v_j$. By part 1 of Lemma \ref{modified_main_lemma},
\begin{eqnarray*}
\left|\sum_{i\leq j} 2^{-i-j} u_i^T A_s'(\omega) v_j\right|  & \leq & c_1\left(\max\left(\sqrt{\lambda \log(d)}, \sqrt{d}\right)\right) \sum_{i=0}^{\infty} |S(u_i)| 2^{-2i} + \left(\frac{\lambda}{c_2} + c_3\sqrt{d}\right) \sum_{j=0}^{\infty} |S(v_j)| 2^{-2j},\\
\left|\sum_{i<j} 2^{-i-j} v_i^T A_s'(\omega) u_j\right| & \leq & c_1\left(\max\left(\sqrt{\lambda \log(d)}, \sqrt{d}\right)\right) \sum_{i=0}^{\infty} |S(v_i)| 2^{-2i} + \left(\frac{\lambda}{c_2} + c_3\sqrt{d}\right) \sum_{j=0}^{\infty} |S(u_j)| 2^{-2j}.
\end{eqnarray*}
Combining the above two we get
\begin{eqnarray*}
\left|a'^T A_s'(\omega) b'\right| 
& = &\left|\sum_{i,j=0}^{\infty} 2^{-i-j} u_i^T A_s(\omega) v_j\right|\\
 &\leq & \left(\frac{\lambda}{c_2} + (c_1+c_3)\left(\max\left(\sqrt{\lambda \log(d)}, \sqrt{d}\right)\right)\right)\left(\sum_{i=0}^{\infty} |S(u_i)| 2^{-2i} + \sum_{j=0}^{\infty}
|S(v_j)2^{-2j}\right)\\ 
& \leq & \left(\frac{4\lambda}{c_2} + 4(c_1+c_3)\left(\max\left(\sqrt{\lambda \log(d)}, \sqrt{d}\right)\right)\right)(\|a'\|^2 + \|b'\|^2).
\end{eqnarray*}
Hence, 
\begin{eqnarray*}
\left|a^T A_s'(\omega) b\right| &\leq & |y^T A_s'(\omega) z| \leq \left(\frac{4\lambda}{c_2} +
4(c_1+c_3)\left(\max\left(\sqrt{\lambda \log(d)}, \sqrt{d}\right)\right)\right)\left(\|a'\|^2 + \|b'\|^2\right)\\
& \leq & \left(\frac{16\lambda}{c_2} + 16(c_1+c_3)\left(\max\left(\sqrt{\lambda \log(d)}, \sqrt{d}\right)\right)\right)\left(\|a\|^2 + \|b\|^2\right).
\end{eqnarray*}
Therefore, 
\begin{eqnarray*}
\left|a^T A_s(\omega) b\right| & \leq & |a^T A_s^1(\omega) b| + |a^T A_s^2(\omega) b| \\
& \leq & \left(\frac{32\lambda}{c_2} + 32(c_1+c_3)\left(\max\left(\sqrt{\lambda \log(d)}, \sqrt{d}\right)\right)\right)\left(\|a\|^2 + \|b\|^2\right).
\end{eqnarray*}

\end{proof}

\begin{proof}[Proof of Claim \ref{claim:generalized_signing_bound_3}]
Since, $a$ is a vector of type $M$, $b$ is a vector of type $L$, we have $|S(u_i)| < \frac{n}{d^2}\le |S(v_j)| $ for all non-zero $v_j$. Applying parts 1 and 2 of Lemma \ref{modified_main_lemma}, we get
\begin{eqnarray*}
\left|\sum_{i\leq j} 2^{-i-j} u_i^T A_s'(\omega) v_j\right|  
& \leq & c_1\left(\max\left(\sqrt{\lambda \log(d)}, \sqrt{d}\right)\right) \sum_{i=0}^{\infty} |S(u_i)| 2^{-2i} + \left(\frac{\lambda}{c_2} + c_3 \sqrt{d}\right) \sum_{j=0}^{\infty} |S(v_j)| 2^{-2j},\\
\left|\sum_{i<j} 2^{-i-j} v_i^T A_s'(\omega) u_j\right| 
& \leq & c_4\left(\max\left(\sqrt{\lambda \log(d)}, \sqrt{d}\right)\right) \left(\sum_{i=0}^{\infty}|S(v_i)| 2^{-2i} + \sum_{j=0}^{\infty} |S(u_j)| 2^{-2j}\right).
\end{eqnarray*}
Combining the above two, we get
\begin{eqnarray*}
\left|a'^T A_s'(\omega) b'\right| & = &\left|\sum_{i,j} 2^{-i-j} u_i^T A_s(\omega) v_j\right| \\
& \leq & \frac{\lambda}{c_2}\sum_j |S(v_j)| 2^{-2j} \\
&& \quad \quad +  (c_1+c_3+c_4)\left(\max\left(\sqrt{\lambda \log(d)}, \sqrt{d}\right)\right)\left(\sum_j |S(v_j)|2^{-2j} +
\sum_i |S(u_i)| 2^{-2i}\right) \\ 
&\leq & \frac{4\lambda}{c_2}\|b'\|^2 + 4(c_1+c_3+c_4)\left(\max\left(\sqrt{\lambda \log(d)}, \sqrt{d}\right)\right)\left(\|b'\|^2 + \|a'\|^2\right).
\end{eqnarray*}
Hence,
\begin{eqnarray*}
\left|a^T A_s'(\omega) b\right| &\leq & \left|a'^T A_s'(\omega) b'\right| \leq \frac{4\lambda}{c_2}\|b'\|^2 + 4(c_1+c_3+c_4)\left(\max\left(\sqrt{\lambda \log(d)}, \sqrt{d}\right)\right)\left(\|b'\|^2 + \|a'\|^2\right)\\
& \leq & \frac{16\lambda}{c_2}\|b\|^2 + 16(c_1+c_3+c_4)\left(\max\left(\sqrt{\lambda \log(d)}, \sqrt{d}\right)\right)\left(\|b\|^2 + \|a\|^2\right).
\end{eqnarray*}
Therefore,
\begin{eqnarray*}
\left|a^T A_s(\omega) b\right| & \leq & \left|a^T A_s^1(\omega) b\right| + \left|a^T A_s^2(\omega) b\right| \\
& \leq & \frac{32\lambda}{c_2}\|b\|^2 + 32(c_1+c_3+c_4)\left(\max\left(\sqrt{\lambda \log(d)}, \sqrt{d}\right)\right)\left(\|b\|^2 + \|a\|^2\right).
\end{eqnarray*}
\end{proof}


\bibliographystyle{alpha}
\bibliography{lifts}
 

\end{singlespace}
\end{document}